\documentclass[journal]{IEEEtran}

\usepackage{xcolor,soul,framed} 

\colorlet{shadecolor}{yellow}
\usepackage[pdftex]{graphicx}
\graphicspath{{../pdf/}{../jpeg/}}
\DeclareGraphicsExtensions{.pdf,.jpeg,.png}

\usepackage[cmex10]{amsmath}
\usepackage{array}
\usepackage{mdwmath}
\usepackage{mdwtab}
\usepackage{eqparbox}
\usepackage{url}
\usepackage{amssymb}
\usepackage{amsmath}
\usepackage{booktabs}
\usepackage{multirow}
\usepackage{algorithm}
\usepackage{algorithmic}
\usepackage{makecell}
\graphicspath{{figure/}}
\newcommand{\tensor}[1]{\boldsymbol{\mathcal{#1}}}

\newcommand{\mat}[1]{\mathbf{#1}}
\newcommand{\vect}[1]{\mathbf{#1}}

\newtheorem{theorem}{Theorem}

\newtheorem{proof}{Proof}[section]
\newtheorem{proposition}[theorem]{Proposition}

\begin{document}
    \title{Fast Hyperspectral Image Recovery via Non-iterative Fusion of Dual-Camera Compressive Hyperspectral Imaging}
    \author{Wei He,~\IEEEmembership{Member,~IEEE,}
      Naoto~Yokoya,~\IEEEmembership{Member,~IEEE,}
      and Xin~Yuan,~\IEEEmembership{Senior Member,~IEEE}

\thanks{Manuscript received March 26, 2019, and accepted June 13, 2019. This work was supported by the Japan Society for the Promotion of Science (KAKENHI 19K20308; KAKENHI 18K18067). (\emph{Corresponding author: wei.he@riken.jp})}
\thanks{W. He and N. Yokoya are with the RIKEN Center for Advanced Intelligence Project, RIKEN, 103-0027, Tokyo, Japan (e-mail: wei.he@riken.jp; naoto.yokoya@riken.jp). N. Yokoya is also with the University of Tokyo, Tokyo, Japan. X. Yuan is with Nokia Bell Labs, USA (xyuan@bell-labs.com). \par
Color versions of one or more of the figures in this letter are available online at http://ieeexplore.ieee.org. Digital Object Identifier XXXX
}
}

\markboth{IEEE TRANSACTIONS ON , VOL.~XX, NO.~XX, XX~2020
}{He W. \MakeLowercase{\textit{et al.}}: Hyperspectral Image Reconstruction from CASSI and RGB Measurements}
\maketitle

\begin{abstract}
Coded aperture snapshot spectral imaging (CASSI) is a promising technique to capture the three-dimensional hyperspectral image (HSI) using a single coded two-dimensional (2D) measurement, in which algorithms are used to perform the inverse problem.
Due to the ill-posed nature, various regularizers have been exploited to reconstruct the 3D data from the 2D measurement. Unfortunately, the accuracy and computational complexity are unsatisfied. One feasible solution is to utilize additional information such as the RGB measurement in CASSI. Considering the combined CASSI and RGB measurement, in this paper, we propose a new fusion model for the HSI reconstruction. 
We investigate the spectral low-rank property of HSI composed of a spectral basis and spatial coefficients. Specifically, the RGB measurement is utilized to estimate the coefficients, meanwhile the CASSI measurement is adopted to provide the orthogonal spectral basis. We further propose a patch processing strategy to enhance the spectral low-rank property of HSI. The proposed model neither requires non-local processing or iteration, nor the spectral sensing matrix of the RGB detector. Extensive experiments on both simulated and real HSI dataset demonstrate that our proposed method outperforms previous state-of-the-art not only in quality but also speeds up the reconstruction more than 5000 times.
\end{abstract}

\IEEEpeerreviewmaketitle

\section{Introduction}
Hyperspectral image (HSI), a three-dimensional (3D) spatial-spectral cube, which is rich in spectral information, has been widely used in applications such as face recognition~\cite{HyperfaceTPAMI2003}, remote sensing~\cite{Stein2002}, food surveillance~\cite{Bioucas2012jstars} and so on. Typically, a spectrometer is used, which scans a 1D-line or 2D plane to capture a full 3D image~\cite{Shaw2003}. Though it can achieve a high spectral resolution such as 10 nm or less~\cite{Green1998}, it is time-consuming and unsuitable for dynamic scenes~\cite{Lizhi_2017PAMI}. Therefore, it is necessary to improve the efficiency of hyperspectral imaging for the dynamic applications.

Inspired by compressive sensing~\cite{Donoho}, several compressive spectral imaging (CSI) techniques have been well developed~\cite{descour1995computed,CASSI2008,ford2001large} to improve the imaging efficiency.
Specifically, a single or few 2D snapshot compressed measurements are captured by the CSI system to reconstruct the 3D HSI data-cube using inverse algorithms.
Among these CSI systems, the coded aperture snapshot spectral imager (CASSI)~\cite{CASSI2008} has attracted much attention, which utilizes a coded aperture and one or two dispersive elements to modulate the optical field from a scene. A detector captures a 2D, multiplexed projection of the 3D spatial-spectral data-cube representing the scene~\cite{CASSI2008}. Following this, different variants of CASSI have been developed, such as dual dispersive CASSI (DDCASSI)~\cite{DDCCSI}, single disperser CASSI (SDCASSI)~\cite{CASSI2008}, spatial-spectral encoded compressive hyperspectral imager (SSCSI)~\cite{SSCSI}, dual-coded hyperspectral imager (DCSI)~\cite{DCSI}, colored coded aperture spectral camera imager (CCASSI)~\cite{arguello2014colored} and so on. For more details, we recommend to the review works~\cite{ImagingRev2011,CASSIreV2016,arce2013compressive}.
In a nutshell, CSI is a system composed of a hardware setup plus a reconstruction algorithm.
In this paper, we focus on the algorithm. In particular, we develop an efficient HSI recovery method via non-iterative fusion of CASSI and the RGB measurements.

%

In order to solve the inverse problem in CSI, {\em i.e.}, retrieving the 3D HSI from the coded 2D measurement, various regularized methods, including sparse representation~\cite{DCSI,SSCSI,RGBRecovery,SIAM2013,zhang2016dictionary}, total variation (TV)~\cite{dualCam2015,yuan2016generalized}, and non-local low-rank regularization~\cite{Yuan_PAMI_2019,Lizhi_2017PAMI,Zhang_2019_ICCV_Ten} have been developed.
Recently, deep learning methods based on the convolutional neural network (CNN) have also been utilized to explore the prior knowledge from coded image to the reconstructed HSI~\cite{Zhang_2019_ICCV,Miao_2019_ICCV}.
However, the quality of the reconstructed HSI is still limited, due to the fact that CASSI is \textit{too} undersampled by compressing HSI from tens to hundreds of spectral bands to one single compressed measurement. In addition, CNN based methods~\cite{Zhang_2019_ICCV,Miao_2019_ICCV} depend on the quality of the training samples, and lose the flexibility~\cite{Yuan_2020}.
In short, the main factors of the reconstruction algorithms for CASSI are the \textit{accuracy, speed and flexibility}~\cite{Yuan_2020}.
However, due to the severe ill-posed nature of CASSI, existing algorithms cannot meet all these three criteria.
One solution from the hardware side to alleviate this challenge is to capture complementary measurements such as panchromatic or RGB images with CASSI~\cite{side_2015,Zhang_2019_ICCV_Ten,CASSI_RGB}.
The related reconstruction algorithms for these hybrid systems~\cite{dualCam2015,Lizhi_2017PAMI,Zhang_2019_ICCV_Ten,side_2015,CASSI_RGB} usually regard the complementary measurements as an addition side information to boost the quality of reconstructed HSI. In particular, these algorithm compose a larger sensing matrix by stacking the complementary sensing matrix and CASSI sensing matrix. That is to say, the computational algorithms to reconstruct HSIs from CASSI and complementary measurements will cost much more time compared to the reconstruction of single CASSI.

Bearing the above concerns in mind, in this paper, we aim to develop an efficient HSI reconstruction algorithm from CASSI and complementary measurements. Different from the previous works, which utilize iteration methods to update the reconstructed HSIs step-by-step to explore the spatial-spectral priors of the full HSIs~\cite{Yuan_PAMI_2019,Lizhi_2017PAMI,Zhang_2019_ICCV_Ten,CASSI_RGB}, our proposed method is inspired by the fact that HSIs are assumed to underlie a low dimensional spectral subspace, which has been widely used in different applications~\cite{BioucasTGRS2008,he2018non,yokoya2012coupled}.
In detail, by taking the advantages of both measurements in the hybrid system, we propose to compute the spatial coefficients from the RGB measurement, and optimize the orthogonal spectral basis from the CASSI measurement, and finally reconstruct the HSI from the product of these two components. We depict the proposed method in Fig.~\ref{fig:fuse}. To make the algorithm efficient, we further propose a non-iterative optimization, which will be described in Section~\ref{sec:fusion}. 
In addition, to reduce the rank of the spatial coefficients, we segment the HSI into spatially overlapping patches, and process each patch separately.
\subsection{Contributions of This Paper}
The main advantages of the proposed method compared with the previous algorithms~\cite{dualCam2015,Lizhi_2017PAMI,Zhang_2019_ICCV_Ten,Zhang_2019_ICCV,Zhao_2019_CVPR,side_2015,CASSI_RGB} can be summarized as follows:

\begin{itemize}[]
\item \textit{Fast:} our proposed two-stage fusion method of CASSI and RGB measurements, does not require regularization or iteration. Therefore, it is efficient to finish the HSI reconstruction within seconds, which is $5000\times$ faster compared to DeSCI~\cite{Yuan_PAMI_2019} and DLTR~\cite{Zhang_2019_ICCV_Ten}.

\item \textit{Flexible:} our model does not need to know the spectral response (sensing matrix) of the RGB detector in advance. What we need is the alignment of CASSI and RGB measurements. It is thus flexible to apply our model to different applications.

\item \textit{High accuracy:} the proposed method is evaluated on extensive simulated and real data experiments. It is reported to achieve the best results compared to other state-of-the-art methods~\cite{Yuan_PAMI_2019,Zhang_2019_ICCV_Ten}.
\end{itemize}

\subsection{Paper Organization and Notations}
The remainder of this paper is organized as follows. Section II introduces the related works of computational imaging reconstruction. Section III introduces the dual-camera compressive hyperspectral imaging system, and presents the proposed fusion model and related solutions. Section IV illustrates the experimental results and Section V presents the detailed discussion of our method, which is followed by the conclusions in Section VI.

In this paper, tensors of order $3$ are denoted by boldface Euler script letters, e.g., $\tensor{X}\in\mathbb{R}^{M\times N \times B}$. Scalars are denoted by normal lowercase letters or uppercase letters, e.g., $x, X \in\mathbb{R}$. $X(i,j,k)$ denotes the element of tensor $\tensor{X}$ in position $(i,j,k)$. Vectors are denoted by boldface lowercase letters, e.g., $\vect{x}\in\mathbb{R}^{M}$. Matrices are denoted by boldface capital letters, e.g., $\mat{X}\in\mathbb{R}^{M\times N}$. The mode-$n$ unfolding \cite{kolda2009tensor} of tensor $\tensor{X}  \in\mathbb{R}^{I\times J \times K}$ is denoted by $\mat{X}_{(n)}\in\mathbb{R}^{B \times  {M N}}$.
The definition of mode-$n$ unfolding is denoted as $\text{fold}_n(\cdot)$, $i.e.,$ for a tensor $\tensor{X}$, we have $\text{fold}_n(\mat{X}_{(n)})=\tensor{X}$.

 \begin{figure}[!t]
  \centering
  \includegraphics[width=1 \linewidth]{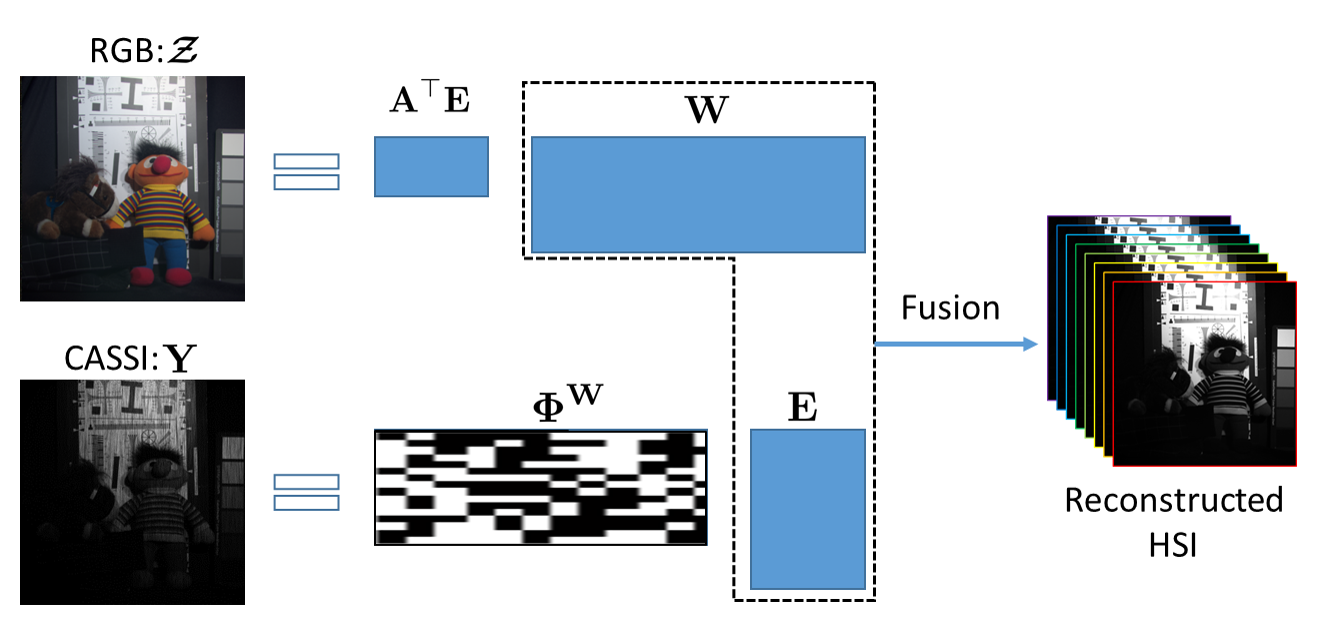}
  \caption{Illustration showing the proposed fusion model.}
  \label{fig:fuse}
  \vspace{-5pt}
\end{figure}

\section{Related Works}
In this section, we briefly introduce the related  reconstruction algorithms from CASSI measurements. \par

To reconstruct HSI from the coded image from CASSI measurements, a two-step iterative shrinkage/thresholding (TwIST) algorithm was firstly utilized to undertake the reconstruction task~\cite{dualCam2015,Sep_imging2013}. As we know, the HSI reconstruction for the highly compressed CASSI measurements is an under-estimated problem. Furthermore, the sensing matrix in CASSI is specifically designed and brings challenge for the reconstruction of HSI~\cite{Yuan_PAMI_2019}. Therefore, different kinds of HSI regularization are introduced into the reconstruction framework to increase the quality of reconstructed HSI. At the beginning, sparse regularization with TI-haar basis~\cite{DDCCSI}, wavelet basis~\cite{CASSI2008} and learned over-complete dictionary~\cite{DCSI,SSCSI,RGBRecovery,SIAM2013,zhang2016dictionary} have been utilized to explore the spatial-spectral sparsity of HSI. To exploit the spatial smoothness of the HSI, total variation (TV) has been utilized into the reconstruction framework~\cite{dualCam2015,yuan2016generalized}. In addition, low-rank matrix/tensor decomposition has also been introduced to explore the spatial-spectral low-rank property~\cite{golbabaee2012joint,wang2017compressive,jiang2018efficient,YingFu2016} and non-local spatial low-rank property of HSI~\cite{Yuan_PAMI_2019,Lizhi_2017PAMI,xue2019nonlocal,Zhang_2019_ICCV_Ten}. As the development of different state-of-the-art HSI regularizers, although the quality of reconstructed HSIs has increased significantly, they also meet new problems. As reported in~\cite{Yuan_PAMI_2019}, the proposed non-local regularized method DeSCI costs more than $3$ hours to reconstruct a HSI of size $512\times 512 \times 31$, preventing the real-time acquisition of HSI. Different regularizers are suitable for different scenario of HSI, which also result in another problem that it is very hard to choose the optimal regularizer for the reconstruction of HSI. \par

Very recently, deep learning has also been utilized to the reconstruction of CASSI~\cite{meng2020end}. \cite{AE_SC_recovery} proposed to use convolutional auto-encoder (AE) to learn a nonlinear sparse representation for CASSI. \cite{Zhao_2019_CVPR} introduced the CNN to train the reconstruction model, and then tested on the coded image. After that, different CNN architectures have been introduced to the reconstruction of coded image via CASSI~\cite{Zhang_2019_ICCV,Miao_2019_ICCV}. The CNN based methods utilize numerous external dataset to train the model, and then apply the trained model to the testing stage. Frankly speaking, the training stage is time-consuming, however the test stage is fast. From this perspective, CNN based methods can alleviate the problems in the previous regularizer based methods, and learn the prior knowledge from the training dataset. However, these end-to-end deep learning methods lose the flexibility~\cite{Yuan_2020}. Since each trained model is allocated to one specific sensing matrix, as the sensing matrix changes, these methods need to train a new model, which is again time-consuming. Furthermore, CNN based methods are out of work when the training samples are insufficient. \par

Although regularization based methods improve the quality of the reconstructed HSI from CASSI measurements, the performance faces the upper bound due the limited measurements. The complex regularizers further bring the heavy computation burden for the reconstruction task. Therefore, some researchers try to reconstruct the HSI from much more measurements, \textit{i.e.,} CASSI measurements with complementary measurements (panchromatic, or RGB). Several improved imagers have been proposed, including multi-frame CASSI~\cite{CASSIreV2016}, dual-camera CASSI with a panchromatic measurement~\cite{dualCam2015,Lizhi_2017PAMI,Zhang_2019_ICCV_Ten}, and dual-camera CASSI with an RGB measurement~\cite{side_2015,CASSI_RGB}. However, the previous works~\cite{dualCam2015,Lizhi_2017PAMI,Zhang_2019_ICCV_Ten,side_2015,CASSI_RGB} regard the complementary measurements as an addition to CASSI, and compose a larger sensing matrix by stacking the complementary sensing matrix and CASSI sensing matrix, which leads to even longer running time.
To reconstruct HSI from CASSI and the complementary measurements, various regularizers have been investigated to restrict full HSI, $i.e.,$ TV~\cite{dualCam2015,CASSI_RGB}, non-local sparse representation~\cite{Lizhi_2017PAMI}, and non-local low-rank matrix/tensor factorization~\cite{Yuan_PAMI_2019,Zhang_2019_ICCV_Ten}.

Compared to CASSI reconstruction, the HSI reconstruction from CASSI and the complementary measurements can achieve higher accuracy, since the additional measurements of complementary image. However, during the optimization, the new sensing matrix (by stacking the complementary sensing matrix and CASSI sensing matrix) has also increased to several times larger than that of CASSI, resulting in the unacceptable time cost of the HSI reconstruction. Furthermore, the optimization needs to know the complementary sensing matrix in advance, which requires additional cross-sensor calibration in the real applications. \par


 \begin{figure}[!t]
  \centering
  \includegraphics[width=1 \linewidth]{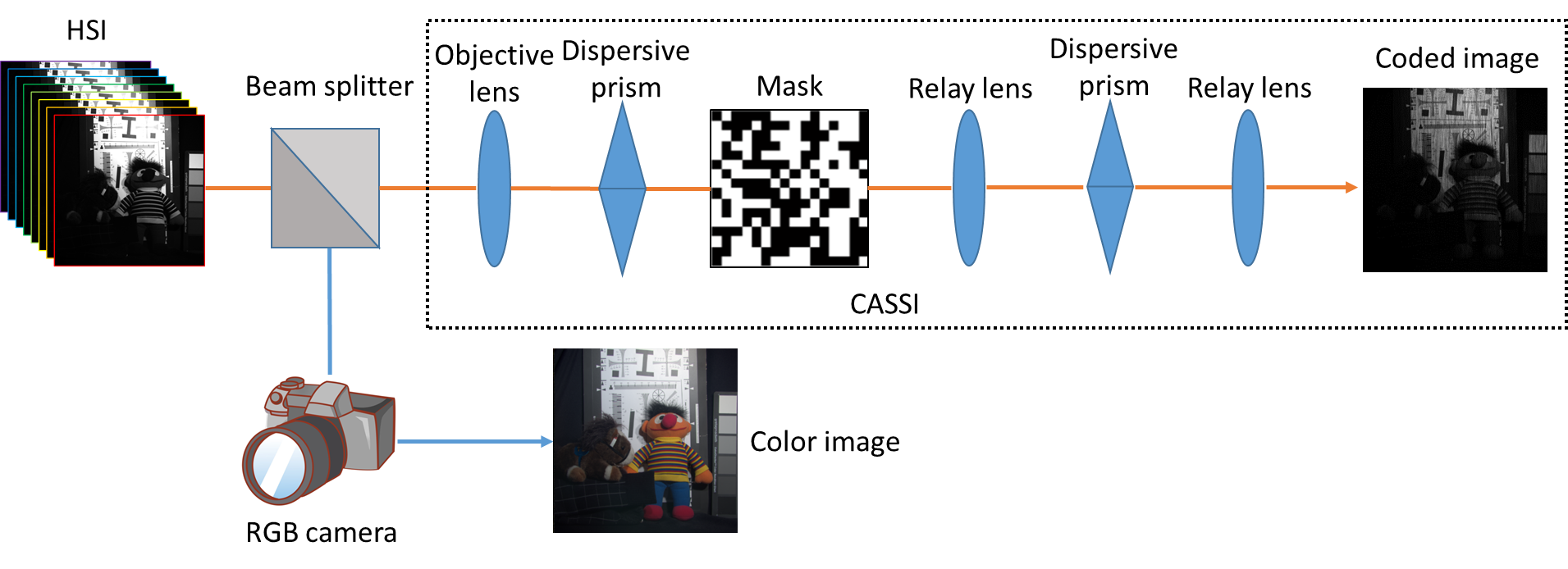}
  \caption{Illustration showing the DCCHI system.}
  \label{fig:ill}
  \vspace{-10pt}
\end{figure}

\section{Proposed Fusion Method}
In this section, we firstly introduce the dual-camera compressive hyperspectral imaging (DCCHI) system. Subsequently, we present the proposed fusion model of CASSI and RGB measurements with  non-iterative optimization strategy. Finally, we analyze the advantages of the proposed model compared to the previous works.
%

\subsection{Dual-camera Compressive Hyperspectral Imaging}
\begin{figure}[!t]
  \centering
  \includegraphics[width=1 \linewidth]{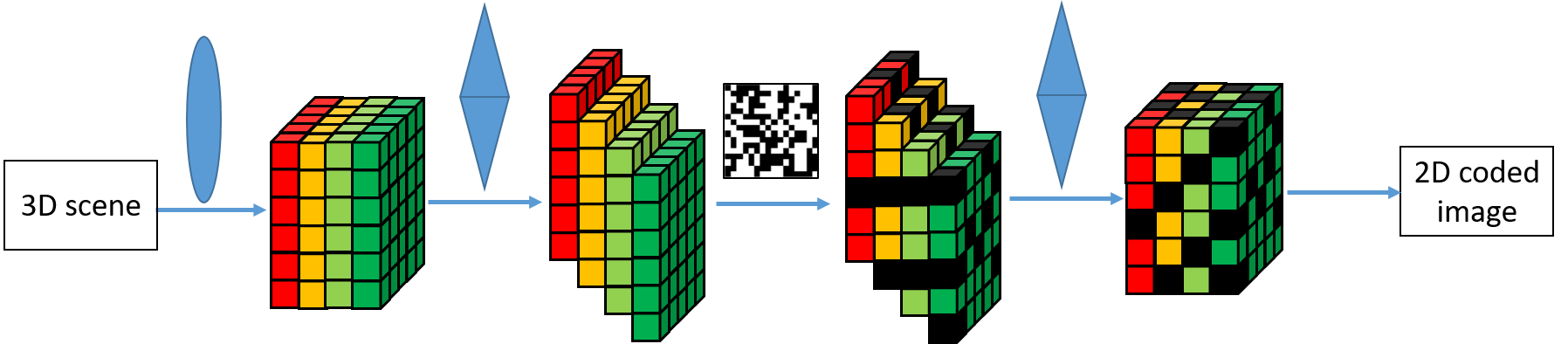}
  \caption{Illustration of CASSI from 3D scene to 2D coded image.}
  \label{fig:ill_cassi}
  \vspace{-10pt}
\end{figure}

The dual-camera compressive hyperspectral imaging system was firstly proposed by~\cite{dualCam2015,side_2015}. As illustrated in Fig. \ref{fig:ill}, the incident light firstly goes through the beam splitter, and is equally partitioned to two directions. The first partition of the incident light is captured by the CASSI, which is composed of objective lens, dispersive prism, mask, relay lens, dispersive prism, and finally the light is sensed by a gray-scale camera to a coded image. We adopt $\tensor{X}\in\mathbb{R}^{M\times N \times B}$ to represent the original HSI, with $M,N$ denoting the two spatial size and $B$ denoting the spectral channels. Therefore, the compressive measurement at position $(i,j)$ via CASSI can be represented as
\begin{equation}
\label{eq:CASSI}
Y(i,j) = \sum_{k=1}^{B} X(i,j,k) \odot C(i,j,k),
\end{equation}
where $\mat{Y} \in \mathbb{R}^{M\times N}$ is the coded image, $\odot$ is the Hadamard (element-wise) product, and $\tensor{C}\in\mathbb{R}^{M\times N \times B}$ denotes the mask. The illustration of CASSI from 3D scene to 2D coded image is presented in Fig~\ref{fig:ill_cassi}. As a linear transformation, \eqref{eq:CASSI} can be reformulated as the following~\cite{Yuan_PAMI_2019,wang2019hyperspectral}
\begin{equation}
\label{eq:CASSI_linear}
\vect{y} = \mat{\Phi}^{c} \vect{x},
\end{equation}
where $\mat{\Phi}^{c} \in  \mathbb{R}^{MN\times MNB}$ is the sensing matrix from $\tensor{C}$, $\vect{y}, \vect{x}$ are the vectorization of $\mat{Y}, \tensor{X}$ respectively. \par

Recalling Fig. \ref{fig:ill}, the second partition of the incident light is captured by the RGB camera and the obtained measurements for each position $(i,j)$  is
\begin{equation}
\label{eq:RGB}
\mat{Z}_{ij:} = \mat{A}^{\top}\mat{X}_{ij:}  .
\end{equation}
Here, $\tensor{Z} \in \mathbb{R}^{M\times N \times 3}$ is the measured RGB image, and $\mat{A}\in \mathbb{R}^{B \times 3}$ is the spectral response function of RGB detector~\cite{yokoya2012coupled}. $\mat{Z}_{ij:}\in \mathbb{R}^{B}$ and $\mat{X}_{ij:}\in \mathbb{R}^{3}$ represent the vectorization of $\mat{Z}(i,j,:)$ and $\mat{X}(i,j,:)$, respectively. Similar to \eqref{eq:CASSI_linear}, \eqref{eq:RGB} can be also reformulated as the linear case
\begin{equation}
\label{eq:RGB_linear}
\vect{z} = \mat{\Phi}^{r} \vect{x},
\end{equation}
where $\mat{\Phi}^{r}\in\mathbb{R}^{3MN\times MNB}$ is the spectral sensing matrix from $\mat{A}$, $\vect{z}$ is the vectorization of $\tensor{Z}$.\par

\subsection{Proposed Observation Model}
To reconstruct the HSI from the CASSI and RGB measurements, previous works~\cite{dualCam2015,Lizhi_2017PAMI,Zhang_2019_ICCV_Ten,Zhang_2019_ICCV,side_2015,CASSI_RGB} stack $\vect{y}$ and $\vect{z}$, meanwhile $\mat{\Phi}^{c}$ and $\mat{\Phi}^{r}$ from \eqref{eq:CASSI_linear} and \eqref{eq:RGB_linear} to build the following framework
\begin{equation}
\label{eq:CASSI_RGB_pre}
\left[\begin{array}{c}\vect{y} \\ \vect{z}\end{array}\right] = \left[\begin{array}{c}\mat{\Phi}^{c}\\
\mat{\Phi}^{r} \end{array} \right] \vect{x}.
\end{equation}
Therefore, \eqref{eq:CASSI_RGB_pre} can be regarded as an extension framework as that of CASSI reconstruction~\eqref{eq:CASSI}. The optimization of \eqref{eq:CASSI_RGB_pre} with proper regularizers~\cite{dualCam2015,CASSI_RGB,Lizhi_2017PAMI,Yuan_PAMI_2019,Zhang_2019_ICCV_Ten} always cost huge computational time.

To avoid the huge computation, different from \eqref{eq:CASSI_RGB_pre}, we assume that {\em the HSIs can be decomposed to the orthogonal spectral basis and related spatial coefficients}. We separately compute the coefficients from the RGB measurement, and the orthogonal spectral basis from the CASSI measurement, and finally fuse the two elements to reconstruct the HSIs.
In detail, HSIs are assumed to lie in an approximate low-dimensional spectral subspace.
Specifically, the spectral low-rank representation of HSIs can be formulated as
\begin{equation}
\label{eq:low-rank}
\mat{X} = \mat{E} \mat{W},
\end{equation}
where we define $\mat{X}:=\mat{X}_{(3)}$ without losing generality, $\mat{E}\in \mathbb{R}^{B\times k}$ and  $\mat{W}\in \mathbb{R}^{k\times MN}$ are the orthogonal spectral basis and coefficients, respectively. $k$ is the rank of $\mat{X}$.


Our major contribution is to embed \eqref{eq:low-rank} into \eqref{eq:RGB} and \eqref{eq:CASSI} to obtain a new observation model from CASSI and RGB measurements
\begin{eqnarray}
\mat{Z} &=& (\mat{A}^{\top}\mat{E}) \mat{W},
\label{eq:fusion1} \\
\vect{y} &=& \mat{\Phi}^{\mat{W}} \vect{e}, \label{eq:fusion2}
\end{eqnarray}
where we again define $\mat{Z}:=\mat{Z}_{(3)}$. $\mat{\Phi}^{\mat{W}}$ is the composition of $\tensor{C}$ and $\mat{W}$, and $\vect{e}$ is the vectorization of $\mat{E}$. We detailed introduce how to obtain $\mat{\Phi}^{\mat{W}}$ from known $\tensor{C}$ and $\mat{W}$.
From \eqref{eq:CASSI}, we can observe that the sensor mask $\tensor{C}$ of CASSI measures the HSI $\tensor{X}$ pixel by pixel. Therefore, \eqref{eq:CASSI} can be reformulated as~\cite{GolbabaeeTIP}
\begin{align}
\label{eq:CASSI_pix}
Y(i,j) &= \mat{C}_{ij:}^{\top} \mat{E}\mat{W}_{ij:} \\ \notag
       &= \{\mat{W}_{ij:}^{\top} \otimes \mat{C}_{ij:}^{\top}\} \vect{e},
\end{align}
where $\mat{C}_{ij:}$ represents the vectorization of $\mat{C}(i,j,:)$, $\mat{W}_{ij:}=\mat{W}(:,i+(j-1)\times N)$, and $\otimes$ is the Kronecker operator. By stacking the measurements of whole pixels together, we can obtain $\mat{\Phi}^{\mat{W}} \in \mathbb{R}^{MN\times kB}$ in \eqref{eq:fusion2} as
\begin{equation}
\label{eq:phi}
\mat{\Phi}^{\mat{W}} = [\mat{W}_{11:}^{\top} \otimes \mat{C}_{11:}^{\top};\ldots;\mat{W}_{MN:}^{\top} \otimes \mat{C}_{MN:}^{\top}].
\end{equation}

The objective of~\eqref{eq:fusion1},~\eqref{eq:fusion2} is to reconstruct the $\tensor{X}$ from the measurements $\mat{Y}$ and $\tensor{Z}$ via CASSI and RGB camera, respectively. Using this new observation model, we can efficiently reconstruct the HSIs using non-iterative fusion method.

\subsection{Proposed Fusion Model}
\label{sec:fusion}
According to the observation model \eqref{eq:fusion1},~\eqref{eq:fusion2}, we try to estimate $\mat{W}$ and $\mat{E}$ separately, and then reconstruct the HSI. As illustrated in Fig. \ref{fig:fuse}, we optimize the coefficients $\mat{W}$ from the RGB measurements $\tensor{Z}$, and then obtain the orthogonal spectral basis $\mat{E}$ from CASSI measurements $\mat{Y}$. Finally, we fuse $\mat{E}$ and $\mat{W}$ to reconstruct the HSI. The proposed fusion model can be formulated as

\begin{eqnarray}
\mat{W} &=& \underset{\mat{W}}{\arg\min} \Vert \mat{Z}- (\mat{A}^{\top}\mat{E}) \mat{W} \Vert_F^2,\label{eq:fusion11} \\
\vect{e} &=& \underset{\vect{e}}{\arg\min} \Vert \vect{y} - \mat{\Phi}^{\mat{W}} \vect{e} \Vert_F^2. \label{eq:fusion22}
\end{eqnarray}

From \eqref{eq:fusion11}, we can easily optimize $\mat{W}$ via singular value decomposition (SVD)~\cite{he2018non}. Subsequently, we can adopt \eqref{eq:phi} to compose $\mat{\Phi}^{\mat{W}}\in \mathbb{R}^{MN \times kB}$. Since $MN \gg kB$, $\mat{\Phi}^{\mat{W}}$ can be regarded as a full-column rank matrix. Therefore, the optimization of $\vect{e}$ can be deduced as
\begin{equation}
\label{eq:opt22}
\vect{e} = \left[(\mat{\Phi}^{\mat{W}})^{\top}(\mat{\Phi}^{\mat{W}})\right]^{-1}(\mat{\Phi}^{\mat{W}})^{\top} \vect{y}.
\end{equation}
Then, we reshape $\vect{e}$ to the matrix version $\mat{E}$, and reconstruct HSI via $\mat{X} = \mat{E} \mat{W}$. The final 3D HSI can be obtained via the folding operator $\tensor{X} = \text{fold}_3(\mat{X})$. \par

Compared to previous works~\cite{Yuan_PAMI_2019,Lizhi_2017PAMI,Zhang_2019_ICCV_Ten,CASSI_RGB} which try to reconstruct the full HSIs, our proposed observation model \eqref{eq:fusion1}, \eqref{eq:fusion2} is to decompose the original $\tensor{X}$ into two very low-dimensional components, and propose to reconstruct the low-dimensional components instead of high dimensional $\tensor{X}$. The spectral low-rank property has been embedded into our observation model, and the ill condition of HSI reconstruction is significantly alleviated. This advantage of our proposed observation model assures the success of non-iterative optimization with no regularization, further saves huge computational time. Furthermore, different from precious works~\eqref{eq:CASSI_RGB_pre}, the spectral sensing matrix of RGB detector in our model is absorbed into the estimation of orthogonal spectral basis, and we do not need to know the spectral sensing matrix in advance.

\subsection{Analysis and Improvements}
\label{Analysis1}
\noindent
\textbf{Fusion model choice:}
From the proposed observation model \eqref{eq:fusion1},~\eqref{eq:fusion2}, we can formulate a generalized fusion model as follows
\
\begin{align}
\label{eq:general}
\{\mat{E},\mat{W}\}
\notag
= \underset{\mat{E},\mat{W}}{\arg\min} &\Vert \mat{Z}- (\mat{A}^{\top}\mat{E}) \mat{W} \Vert_F^2 + \Vert \vect{y} - \mat{\Phi}^{\mat{W}} \vect{e} \Vert_F^2 \\
&+ \lambda_1 \Vert \mat{W} \Vert_{reg1} + \lambda_2 \Vert \mat{E} \Vert_{reg2}
\end{align}
to jointly update $\mat{E}$ and $\mat{W}$. Here, $\Vert \mat{W} \Vert_{reg1}$ and $\Vert \mat{E} \Vert_{reg2}$ mean the regularizers on $\mat{W}$ and $\mat{E}$, separately; $\lambda_1$ and $\lambda_2$ are the parameters to balance the contributions from each regularizers. Typically, TV~\cite{dualCam2015,CASSI_RGB}, non-local sparse representation~\cite{Lizhi_2017PAMI}, and non-local low-rank matrix/tensor factorization~\cite{Yuan_PAMI_2019,he2018non} can be utilized to regularize $\mat{W}$ and spectral smoothness can be utilized to restrict $\mat{E}$. However, we utilize the proposed fusion model \eqref{eq:fusion11}, \eqref{eq:fusion22} rather than \eqref{eq:general} to reconstruct the HSI due to the following reasons. $i$) Our proposed model do not require the additional regularizers on $\mat{W}$ and $\mat{E}$, which significantly reduces the computation time; meanwhile we don't need to know the spectral sensing matrix $\mat{A}$ in advance, which increases the applicability in different settings. $ii$) As reported in the experimental section, our proposed fusion model has achieved the state-of-the-art reconstruction results compared to existing non-training based methods. $iii$) Theoretical guarantee can be weakly obtained to our proposed fusion model.

\begin{proposition}{(conditioned exact reconstruction)}
\label{pr:cond}
If we have the conditions: 1) the spectral sensing matrix of RGB is of full-column rank; 2) the composed $\mat{\Phi}^{\mat{W}}$ is of full-column rank; and 3) $k\leqslant 3$. Then $\tensor{X}$ is the exact reconstruction from the measurements $\mat{Y}$ and $\tensor{Z}$ via $\text{fold}_3(\mat{\mat{E}\mat{W}})$, with $\mat{E}$ from SVD, and $\mat{W}$ from \eqref{eq:opt22}.
\end{proposition}
The proof is provided in the Appendix. In Proposition~\ref{pr:cond}, condition 1) is related to the design of RGB camera. As reported in \cite{qu2018unsupervised}, the spectral sensing matrix in Nikon D700 is of full-column rank. That is to say, condition 1) is easy to obtain. Furthermore, since $MN \gg kB$, $\mat{\Phi}^{\mat{W}}$ can be regarded as a full-column rank matrix with specific pattern $\mat{\Phi}$. The only problem of condition is that $k\leqslant 3$, which is not the real case in most of HSIs. We will propose an approach to meet this condition and thus an improvement of the fusion algorithm in the next subsection.

\begin{figure}[htbp!]
  \centering
  \includegraphics[width=1 \linewidth]{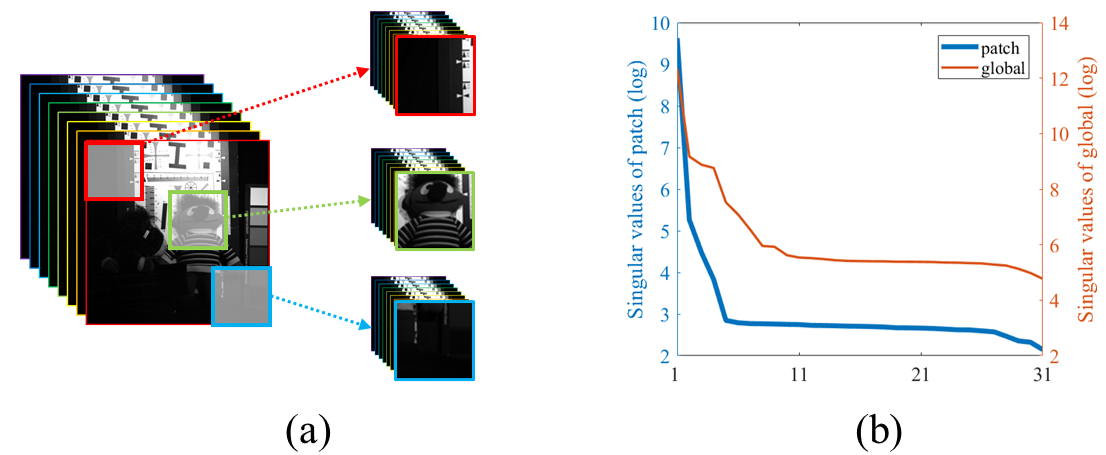}
  \caption{(a) 3D patch extracted from the Toy dataset; (b) singular values in logarithm of reshaped patches V.S global image. We select 100 patches and average the singular values in logarithm.}
  \label{fig:SVD}
\end{figure}
\noindent
\textbf{Patch based fusion algorithm:}
As analyzed before, the condition of $k\leqslant 3$ is unreasonable. Generally speaking, for a HSI scenario, the rank is always larger that $3$~\cite{BioucasTGRS2008,he2018non}. However, from another perspective, the HSIs usually have the piece-wise smoothness property~\cite{Yuan_PAMI_2019,YingFu2016}. That is to say, the pixel signatures from a local spatial patch have a higher probability of being similar. In Fig. \ref{fig:SVD}, we illustrate the comparison of the singular values in logarithm between the extracted patch images and the global Toy image. We choose 100 patch images of size $100 \times 100 \times 31$ and plot the average singular values. From the figure, it can be seen that the average singular values of patch images drop much faster than the ones from the global image, indicating that local patch analysis can significantly increase the spectral low-rank property~\cite{he2015hyperspectral}. To utilize this finding, we propose to segment the HSIs to overlapping patches, and reconstruct each patch separately (and in parallel). Finally, the whole reconstructed patches are composed to the final reconstructed HSI. The proposed patch based fusion algorithm is presented in Algorithm~\ref{alg:pfusion}. What we need to mention is that we should obey $mn > kB$, where $m, n$ are the patch spatial size, to make sure $\mat{\Phi}^{\mat{W}}$ satisfy the full-column rank condition. \par

The patch based fusion can enhance the spectral low-rank property and increase the HSI reconstruction performance, as analyzed in Table~\ref{tab:patchsize}. However, the condition of $k\leqslant 3$ is not always strictly satisfied, as presented in Fig. \ref{fig:SVD}. Furthermore, the patch based fusion also increases the computational time. This means that the proposed patch based fusion algorithm still has room for improvement. Another strategy to alleviate this condition is to explore priors with regularization. However, as we mentioned before, the regularizations will significantly improve the computational burden.  We leave this question, \textit{i.e.,} the condition of $k\leqslant 3$, as future work.

\begin{algorithm}[tp]
\caption{Patch based fusion}
\label{alg:pfusion}
\begin{algorithmic}[1]
	\REQUIRE $\mat{Y}$ measured via CASSI, $\tensor{Z}$ measured via RGB camera, CASSI sensing matrix $\mat{\Phi}$, rank $k$.
	\STATE Initialization: segment $\mat{Y}$, $\tensor{Z}$ and $\mat{\Phi}$ into aligned overlapping patches of size $m \times n$, $m \times n \times 3$, and $m \times n \times B$ respectively.
	\FOR{each aligned patch}
	\STATE {A). Update the coefficients via SVD.}
	\STATE {B). Update the orthogonal spectral basis via \eqref{eq:opt22}.}
	\STATE {C). Reconstruct each patch of size $m \times n \times B$.}
	\ENDFOR	
	\RETURN  Aggregate all the patches into the final $\mathcal{X}$;
\end{algorithmic}
\end{algorithm}

\noindent
\textbf{Orthogonal spectral basis estimation improvement:} The optimization of $\mat{E}$ via \eqref{eq:opt22} simply utilized the measurement $\tensor{C}$, ignoring the RGB sensor matrix $\mat{A}$. Similar to \eqref{eq:CASSI_pix}, the RGB measurement can be also formulated as the pixel-by-pixel version
\begin{align}
\label{eq:RGB_pix}
Z(i,j,k) &= (\mat{C}(:,k))^{\top} \mat{E}\mat{W}_{ij:} \\ \notag
       &= \{\mat{W}_{ij:}^{\top} \otimes (\mat{C}(:,k))^{\top}\} \vect{e}.
\end{align}
So we can obtain $\mat{\Phi}_{RGB}^{\mat{W}}$ in the similar way of \eqref{eq:phi}, and update $\vect{e}$ via
\begin{equation}
\label{eq:fusion22_ext}
\vect{e} = \underset{\vect{e}}{\arg\min} \left\| [\vect{y};\vect{z}] - [\mat{\Phi}^{\mat{W}};\mat{\Phi}_{RGB}^{\mat{W}}] \vect{e} \right\|_F^2
\end{equation}
instead of \eqref{eq:fusion22}. In Section \ref{sec:improvel}, we compare the results obtained by \eqref{eq:fusion22} and \eqref{eq:fusion22_ext}, respectively. As reported, optimizing $\mat{E}$ via \eqref{eq:fusion22_ext} results in the limited improvement of accuracy, unfortunately, increasing almost $4\times$ more computation time. We thus recommend to use \eqref{eq:fusion22} to update the orthogonal spectral basis $\mat{E}$.

\begin{table}[ht]
\footnotesize
\centering
\vspace{-5px}
\caption{The size of the dataset used in the experiments.}
\begin{tabular}{c | c | c |c}
	\hline
	  Image name   & HSI                       & CASSI            & RGB                         \\ \hline
	  CAVE         & 512$\times$512$\times$31  & 512$\times$512   & 512$\times$512$\times$3     \\ \hline
	  ICVL         & 512$\times$512$\times$31  & 512$\times$512   & 512$\times$512$\times$3     \\ \hline
	  RS           & 400$\times$200$\times$128 & 400$\times$200   & 400$\times$200$\times$3     \\ \hline
      Bird         & 1021$\times$703$\times$24 & 1021$\times$703  & 1021$\times$703$\times$3       \\ \hline
\end{tabular}
\label{tab:datasize}
\vspace{-5px}
\end{table}

\begin{table*}[!htbp]
  \centering
  \caption{Quantitative evaluation of CAVE data experiments for different HSI reconstruction methods from CASSI and RGB measurements.}
    \begin{tabular}{cccccccccccc|c}
    \toprule
    CAVE  & method & balloons & beads & cd    & toy   & clay  & cloth & egyptian & face  & beers & food  & Average \\
    \midrule
    \multirow{7}[2]{*}{M-PSNR}
          & TV     & 31.10    & 20.82 & 28.73 & 25.93 & 27.75 & 22.98 & 33.70 & 31.86 & 30.07 & 28.89 & 28.18 \\
          & DeSCI  & 32.14    & 21.76 & 29.04 & 28.70 & 28.15 & 25.30 & 36.10 & 33.49 & 31.28 & 29.78 & 29.57 \\
          & TV-RGB & 36.75    & 26.67 & 31.68 & 29.19 & 36.42 & 25.45 & 37.49 & 35.82 & 35.13 & 34.27 & 32.89 \\
          & DeSCI-RGB & 40.35 & 30.54 & 32.94 & 34.36 & 39.93 & 30.27 & 42.23 & 39.78 & 39.28 & 37.67 & 36.74 \\
          & DLTR   & 41.35	  & 28.66 & 32.26 &	33.12 &	39.68 &	29.00 &	42.13 &	43.63 &	41.29 & 37.82 & 36.89  \\
          & Fusion & 39.50    & 32.30 & 29.25 & 40.02 & 38.94 & 38.34 & 48.40 & 43.10 & 41.01 & 38.86 & 38.97 \\
          & PFusion & \textbf{48.50} & \textbf{32.91} & \textbf{32.95} & \textbf{44.33} & \textbf{49.24} & \textbf{39.28} & \textbf{50.93} & \textbf{46.40} & \textbf{46.49} & \textbf{43.12} & \textbf{43.42} \\
    \midrule
    \multirow{7}[2]{*}{M-SSIM}
          & TV     & 0.952    & 0.659 & 0.922 & 0.880 & 0.845 & 0.660 & 0.948 & 0.956 & 0.946 & 0.852 & 0.862 \\
          & DeSCI  & 0.963    & 0.727 & 0.939 & 0.925 & 0.861 & 0.819 & 0.968 & 0.971 & 0.960 & 0.878 & 0.901 \\
          & TV-RGB & 0.964    & 0.878 & 0.908 & 0.944 & 0.911 & 0.783 & 0.980 & 0.971 & 0.946 & 0.913 & 0.920 \\
          & DeSCI-RGB & 0.991 & 0.942 & 0.962 & 0.981 & 0.954 & 0.923 & 0.994 & 0.990 & 0.991 & 0.956 & 0.968 \\
          & DLTR   & 0.985    & 0.903 &	\textbf{0.975} &	0.971 &	0.973 &	0.900 &	0.981 &	0.992 &	0.992 & 0.977 &	0.965  \\
          & Fusion & 0.985    & 0.933 & 0.916 & 0.968 & 0.934 & 0.975 & 0.983 & 0.981 & 0.995 & 0.952 & 0.962 \\
          & PFusion & \textbf{0.995} & \textbf{0.944} & 0.971 & \textbf{0.989} & \textbf{0.994} & \textbf{0.977} & \textbf{0.996} & \textbf{0.994} & \textbf{0.997} & \textbf{0.980} & \textbf{0.984} \\
    \midrule
    \multirow{7}[2]{*}{MSA}
          & TV     & 10.21    & 25.35 & 13.01 & 12.49 & 17.15 & 16.30 & 14.21 & 10.29 & 4.80  & 16.12 & 13.99 \\
          & DeSCI  & 9.79     & 25.28 & 11.08 & 12.06 & 15.94 & 14.48 & 13.43 & 9.49  & 4.72  & 15.09 & 13.14 \\
          & TV-RGB & 6.63     & 13.78 & 12.40 & 10.36 & 14.24 & 14.22 & 11.33 & 8.83  & 3.71  & 13.40 & 10.89 \\
          & DeSCI-RGB & 4.64  & \textbf{9.96} & 7.44  & 7.23  & 9.40  & 10.16 & \textbf{8.23} & 6.60  & 2.60  & \textbf{9.57} & 7.58 \\
          & DLTR   & 6.24	  & 13.79 &	8.03 &	8.67  &	15.38 &	9.23  &	13.27 &	7.10 &	1.27  &	6.43 &	8.94 \\
          & Fusion & 9.91     & 17.65 & 28.94 & 15.90 & 29.15 & 5.84  & 23.38 & 15.58 & 2.19  & 23.99 & 17.25 \\
          & PFusion  & \textbf{4.54} & 14.59 & \textbf{7.26} & \textbf{6.81} & \textbf{7.99} & \textbf{5.07} & 9.25  & \textbf{6.37} & \textbf{1.21} & 10.77 & \textbf{7.39} \\
    \bottomrule
    \end{tabular}%
  \label{tab:eva1}%
\end{table*}%
  \begin{figure*}[!htp]
\centering
 \begin{minipage}[t]{0.18\textwidth}\centering
   \includegraphics[width=\textwidth]{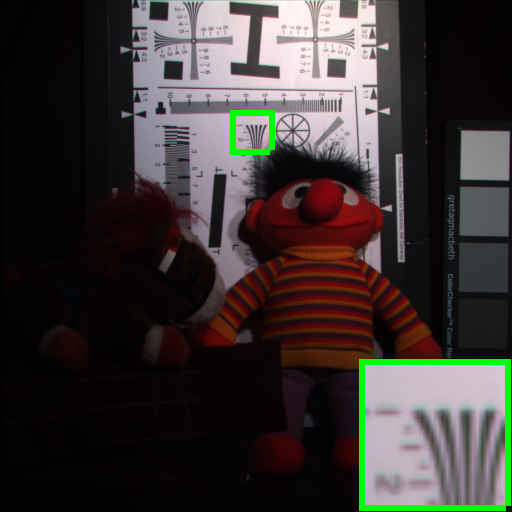}\\(a) Original
 \end{minipage}
 \begin{minipage}[t]{0.18\textwidth}\centering
   \includegraphics[width=\textwidth]{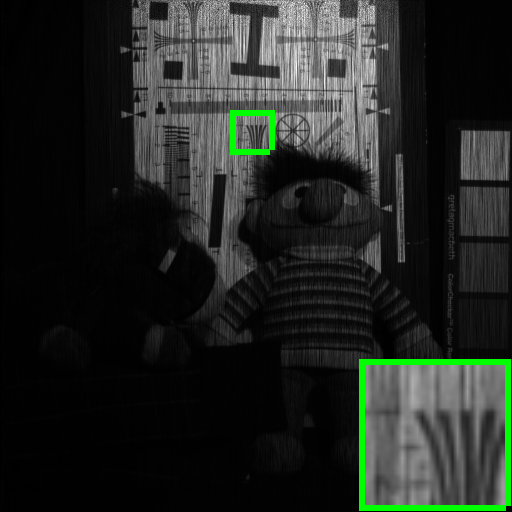}\\(b) CASSI
 \end{minipage}
 \begin{minipage}[t]{0.18\textwidth}\centering
   \includegraphics[width=\textwidth]{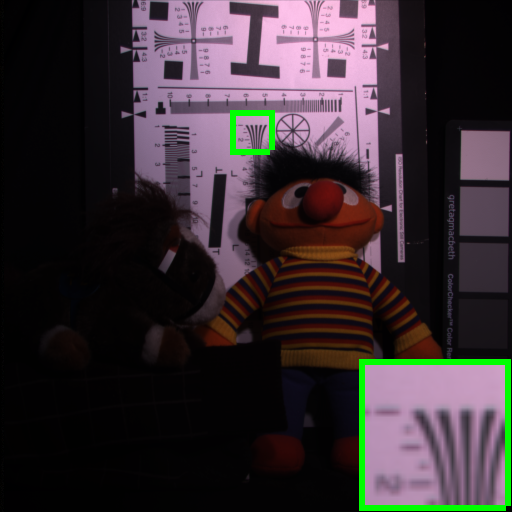}\\(c) RGB
 \end{minipage}
 \begin{minipage}[t]{0.18\textwidth}\centering
   \includegraphics[width=\textwidth]{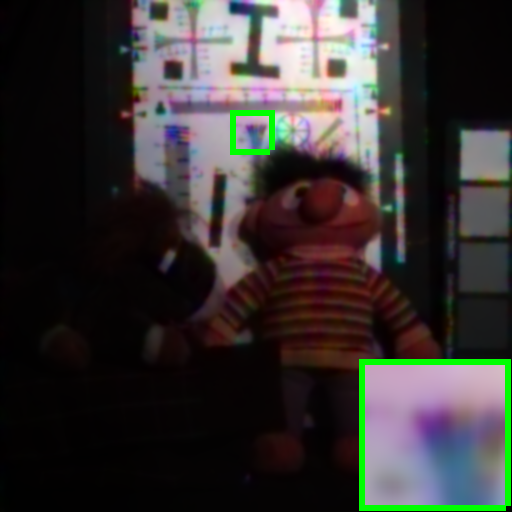}\\(d) TV
 \end{minipage}
 \begin{minipage}[t]{0.18\textwidth}\centering
   \includegraphics[width=\textwidth]{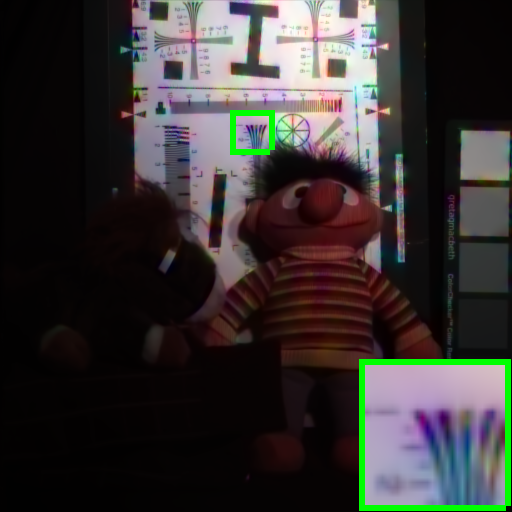}\\(e) DeSCI
 \end{minipage}\\

 \begin{minipage}[t]{0.18\textwidth}\centering
   \includegraphics[width=\textwidth]{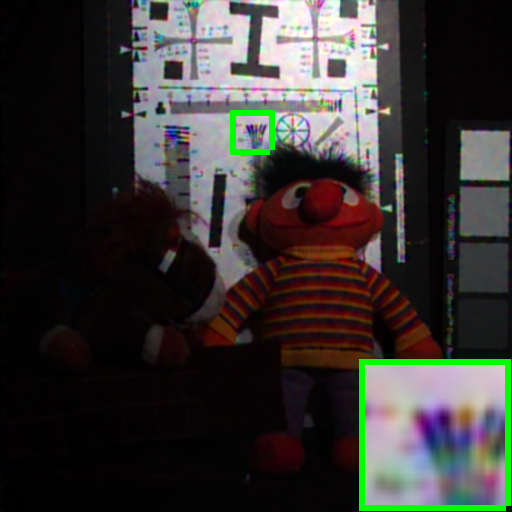}\\(f) TV-RGB
 \end{minipage}
 \begin{minipage}[t]{0.18\textwidth}\centering
   \includegraphics[width=\textwidth]{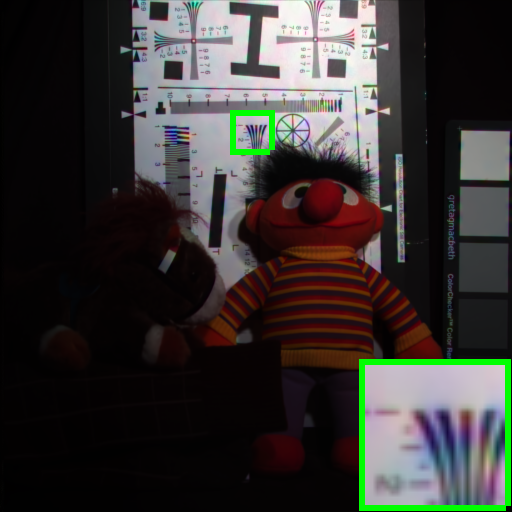}\\(g) DeSCI-RGB
 \end{minipage}
 \begin{minipage}[t]{0.18\textwidth}\centering
   \includegraphics[width=\textwidth]{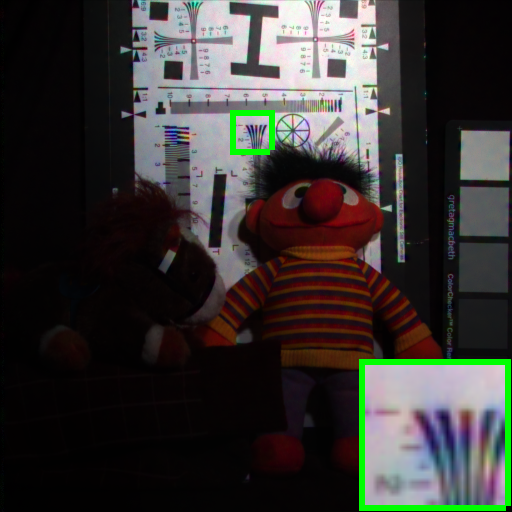}\\(h) DLTR
 \end{minipage}
 \begin{minipage}[t]{0.18\textwidth}\centering
   \includegraphics[width=\textwidth]{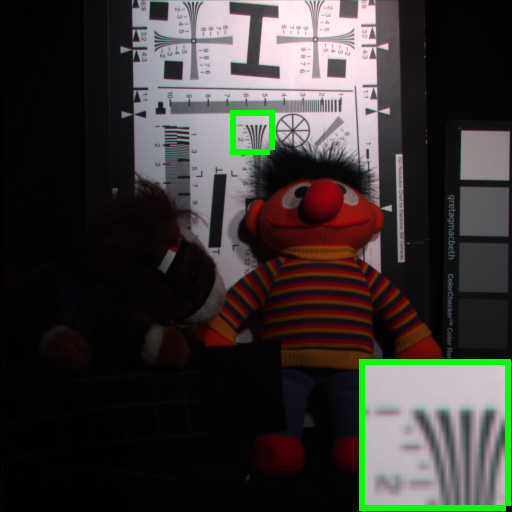}\\(i) Fusion
 \end{minipage}
 \begin{minipage}[t]{0.18\textwidth}\centering
   \includegraphics[width=\textwidth]{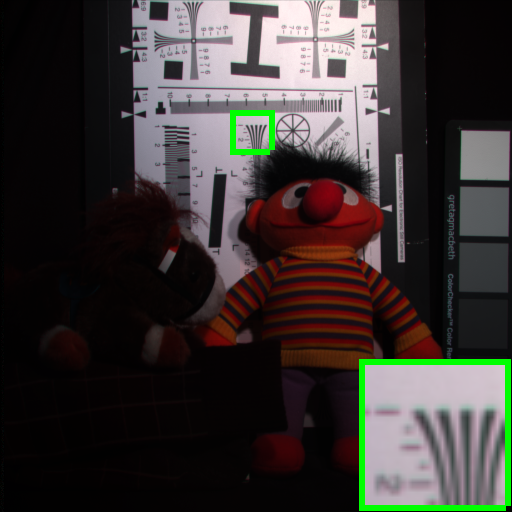}\\(j) PFusion
 \end{minipage}\\
\caption{Reconstructed results of different methods on CAVE-Toy image. The color images are composed of bands 31, 11, 6 for red, green and blue, respectively.}
\label{fig:CAVE_toy}
 \end{figure*}

\section{Experimental Results}
\label{Experiments}
In this section, we adopted simulated and real dataset experiments to demonstrate the advantage of the proposed methods on the reconstruction of CASSI and RGB measurements compared to other state-of-the-art methods. We employ the following methods for comparison:
\textit{CASSI reconstruction methods}, \textit{i.e.}
GAP-TV \cite{yuan2016generalized},
DeSCI \cite{Yuan_PAMI_2019}\footnote{\url{ https://github.com/liuyang12/DeSCI}}
and \textit{CASSI \& RGB reconstruction methods}, \textit{i.e.}
TV-RGB \cite{CASSI_RGB},
DeSCI-RGB\footnote{We implement the algorithm by using~\eqref{eq:CASSI_RGB_pre} instead of \eqref{eq:CASSI_linear}.}
and DLTR \cite{Zhang_2019_ICCV_Ten}\footnote{Thank Dr. S. Zhang for providing the experimental results.}.
The proposed algorithm for the global reconstruction is marked as `Fusion', meanwhile the proposed patch based fusion algorithm is denoted as `PFusion'.

In the simulated experiments, the peak signal-to-noise ratio (PSNR), structural similarity (SSIM) and the mean of the spectral angle (MSA)~\cite{Zhang_2019_ICCV,he2018non} between the reconstructed and original images are used to evaluate the results. Since HSIs have multi spectral bands, we calculate the mean PSNR and mean SSIM of each band, and then average them~\cite{Zhang_2019_ICCV,he2018non}, which are denoted as `M-PSNR' and `M-SSIM', respectively. The whole experiments (except DLTR) are programmed in Matlab R2017b on a laptop with CPU Core i7-8750H 16G memory~\footnote{DLTR was programmed on a laptop with CPU Core i7-6700 64G memory}. TV-RGB and DeSCI-RGB dealing with the global HSIs are out of memory on the laptop. Therefore, as the same of `PFusion', we segment the HSIs to overlapping patches and adopt TV-RGB and DeSCI-RGB to process small patches separately.

\subsection{Simulated Experiments On CAVE}
The first test dateset is CAVE\footnote{\url{http://www1.cs.columbia.edu/CAVE/databases/}}, which contains $32$ HSI scenes. We select the first $10$ images for the experiments. We adopted the CASSI sensing matrix from \cite{Yuan_PAMI_2019} to simulate the CASSI measurements, meanwhile spectral sensing matrix of Nikon D700 camera \cite{qu2018unsupervised} to simulate the RGB measurements. The size of original HSI, CASSI measurements and RGB measurements are presented in Table \ref{tab:datasize}. For our proposed methods, we choose $k=3$ and $m=n=100$.


\begin{figure}[!htp]
 \centering
 \begin{minipage}[t]{0.24\textwidth}\centering
   \includegraphics[width=\textwidth]{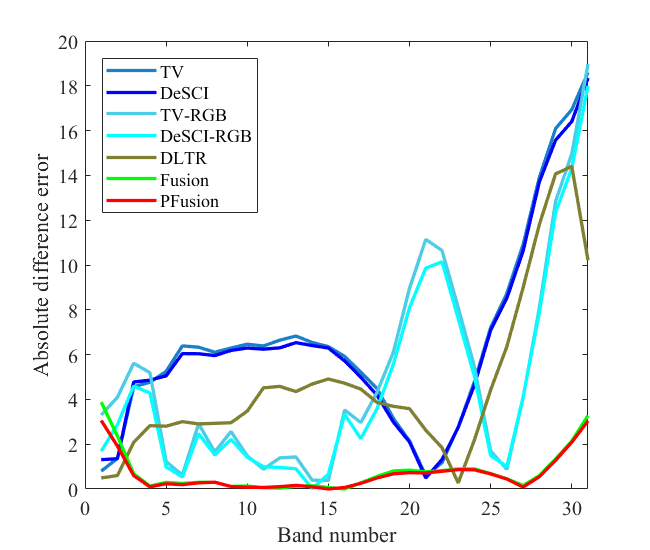}\\(a) beads
 \end{minipage}
 \begin{minipage}[t]{0.24\textwidth}\centering
   \includegraphics[width=\textwidth]{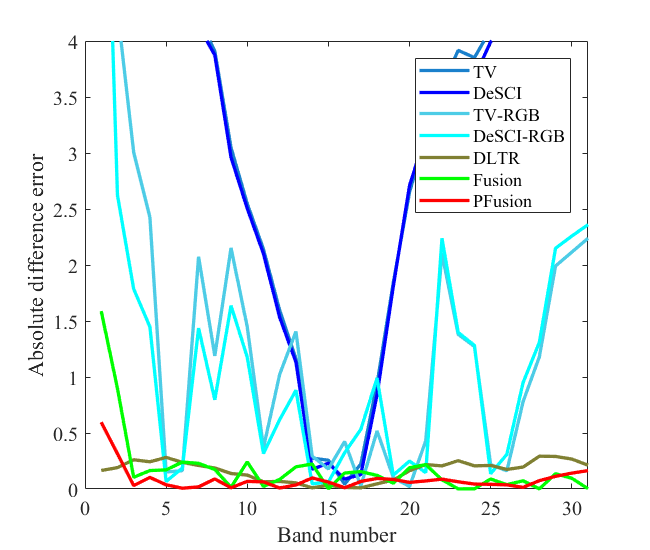}\\(b) food
 \end{minipage}\\
 \caption{The absolute difference errors (smaller is better) between the means of the whole original signatures and the one obtained by different reconstruction methods on CAVE images.}
 \label{fig:CAVE_sig}
 \vspace{-5pt}
\end{figure}

\noindent
\textbf{Quantitative comparison.}
We present the M-PSNR, M-SSIM and MSA values of reconstructed images via different methods in Table \ref{tab:eva1}. Our proposed PFusion method achieves the best values in average of ten different HSI scenes of all these evaluation metrics, conforming the advantage of our proposed method. Compared to CASSI reconstruction methods TV and DeSCI, the reconstruction from CASSI and RGB measurements can significantly improve the quality of the reconstructed image. This benefit is from the additional RGB measurements. DeSCI-RGB and DLTR can achieve satisfied results, indicating the efficiency of non-local low-rank processing. Our proposed method can achieve higher MPSNR values compared to DeSCI-RGB, meanwhile lower MSA values. This is mainly because our proposed Fusion can take full advantage of RGB measurements to enhance the spatial characteristic, unfortunately fails to take the full features of global spectral property within rank $k=3$. Therefore, the spectral distortion of our proposed Fusion is severe. Our proposed PFusion, which processes overlapping patches instead of global image, can significantly improve the reconstructed spectral quality with smaller MSA values. Furthermore, the M-PSNR and M-SSIM values obtained by PFusion are further improved compared to those of Fusion.

\noindent
\textbf{Visual comparison.}
To further illustrate the efficiency of different methods, we present the color images (composed of bands 31, 21 and 6~\cite{he2018non}) of different reconstruction methods on Toy image in Fig.~\ref{fig:CAVE_toy}. From the figure, it can be observed that the RGB measurements are different from the original images composed of bands 31, 11 and 6. TV based method results in the blurred results. DeSCI can produce clearer image, and DeSCI-RGB and DLTR can further improve the results. However, the spectral distortion is also obvious, as some undesired color artifacts appeared in the results of DeSCI-RGB and DLTR. Our proposed Fusion and PFusion can produce the most similar results compared to the original image presented in Fig.~\ref{fig:CAVE_toy}(a). \par

To further check the reconstructed spectrum, we also present the absolute values between the means of the whole original signatures and the one obtained by different reconstruction methods. We choose two images, $i.e.,$ Beads and food for illustration. From Fig.~\ref{fig:CAVE_sig}, we can obtain that Fusion method can achieve similar results compared to DeSCI-RGB and DLTR. Our proposed PFusion can significantly improve the spectral quality compared to Fusion, indicating the significant advantage of our proposed PFusion to process overlapping patches instead of the global image.

\subsection{Simulated Experiments On ICVL}

\begin{table}[!htbp]
  \centering
  \caption{Quantitative evaluation of ICVL data experiments for different HSI reconstruction methods from CASSI and RGB measurements.}
   \setlength{\tabcolsep}{1.5mm}{
    \begin{tabular}{ccccccc|c}
    \toprule
    ICVL  & method    & 4cam  & BGU   & IDS   & Ist   & prk   & Average \\
    \midrule
    \multirow{7}[2]{*}{M-PSNR}
          & TV        & 31.11 & 28.93 & 28.39 & 34.85 & 32.42 & 31.14 \\
          & DeSCI     & 33.17 & 31.38 & 30.46 & 35.44 & 34.24 & 32.94 \\
          & TV-RGB    & 35.20 & 33.80 & 32.67 & 38.34 & 37.67 & 35.54 \\
          & DeSCI-RGB & 40.39 & 40.30 & 37.91 & 39.93 & 43.89 & 40.48 \\
          & DLTR      &41.28  & 39.12 &	36.33 &	39.98 &	39.01 &	38.04  \\
          & Fusion    & 43.65 & 44.91 & 43.19 & 37.13 & 46.95 & 43.17 \\
          & Pfusion   & \textbf{48.76} & \textbf{48.94} & \textbf{45.07} & \textbf{41.91} & \textbf{49.87} & \textbf{46.91} \\
    \midrule
    \multirow{7}[2]{*}{M-SSIM}
          & TV        & 0.912 & 0.858 & 0.912 & 0.967 & 0.907 & 0.911 \\
          & DeSCI     & 0.953 & 0.927 & 0.951 & 0.974 & 0.940 & 0.949 \\
          & TV-RGB    & 0.946 & 0.931 & 0.928 & 0.982 & 0.961 & 0.950 \\
          & DeSCI-RGB & 0.992 & 0.990 & 0.992 & 0.990 & 0.994 & 0.991 \\
          & DLTR      & 0.986 &	0.979 &	0.983 &	0.991 &	0.969 &	0.975  \\
          & Fusion    & 0.993 & 0.992 & 0.993 & 0.985 & 0.993 & 0.991 \\
          & Pfusion   & \textbf{0.995} & \textbf{0.996} & \textbf{0.995} & \textbf{0.992} & \textbf{0.996} & \textbf{0.995} \\
    \midrule
    \multirow{7}[2]{*}{MSA}
          & TV        & 4.36  & 6.50  & 4.34  & 5.08  & 5.84  & 5.23 \\
          & DeSCI     & 3.87  & 5.57  & 3.94  & 4.59  & 4.76  & 4.55 \\
          & TV-RGB    & 3.12  & 4.41  & 2.92  & 2.92  & 3.87  & 3.45 \\
          & DeSCI-RGB & 1.93  & 2.28  & 1.81  & 1.97  & 1.94  & 1.99 \\
          & DLTR      &1.23	  &2.27   &	0.96  &	1.25  &	2.97  &	1.74  \\
          & Fusion    & 1.91  & 2.28  & 0.94  & 5.20  & 2.49  & 2.56 \\
          & Pfusion   & \textbf{0.82} & \textbf{1.16} & \textbf{0.64} & \textbf{1.94} & \textbf{1.21} & \textbf{1.15} \\
    \bottomrule
    \end{tabular}}%
  \label{tab:eva2}%
\end{table}%

  \begin{figure*}[!htp]
\centering
 \begin{minipage}[t]{0.18\textwidth}\centering
   \includegraphics[width=\textwidth]{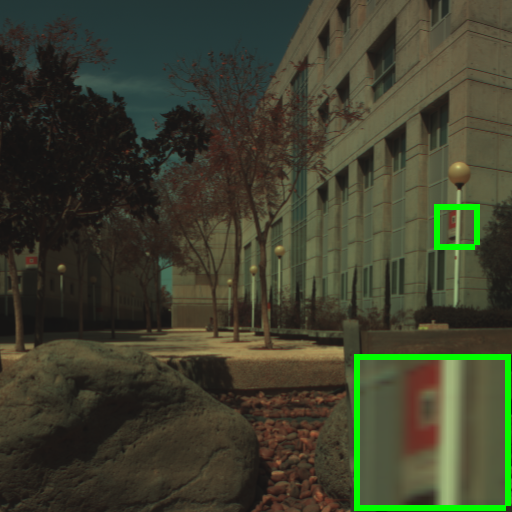}\\(a) Original
 \end{minipage}
 \begin{minipage}[t]{0.18\textwidth}\centering
   \includegraphics[width=\textwidth]{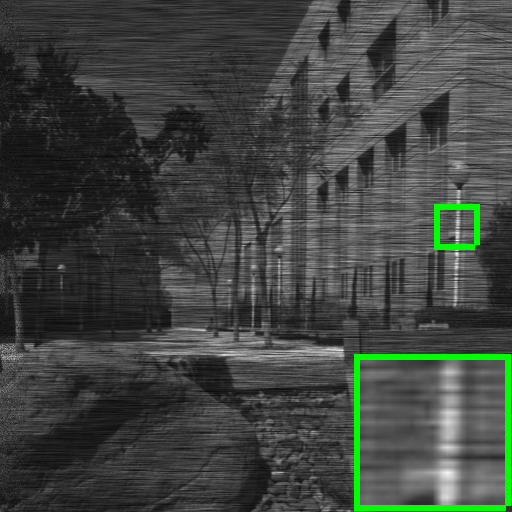}\\(b) CASSI
 \end{minipage}
 \begin{minipage}[t]{0.18\textwidth}\centering
   \includegraphics[width=\textwidth]{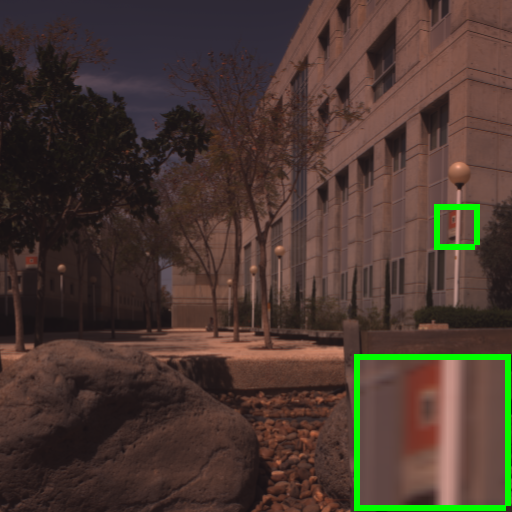}\\(c) RGB
 \end{minipage}
 \begin{minipage}[t]{0.18\textwidth}\centering
   \includegraphics[width=\textwidth]{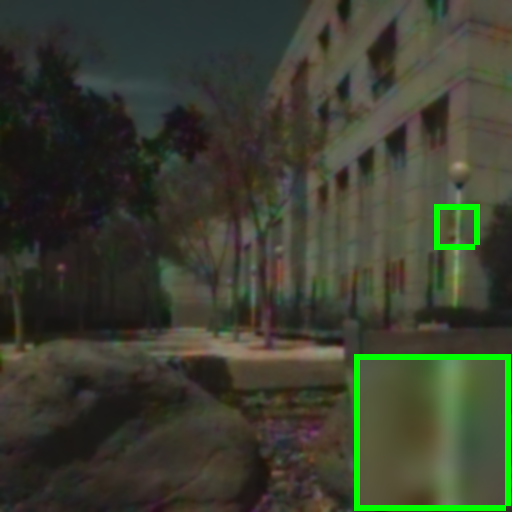}\\(d) TV
 \end{minipage}
 \begin{minipage}[t]{0.18\textwidth}\centering
   \includegraphics[width=\textwidth]{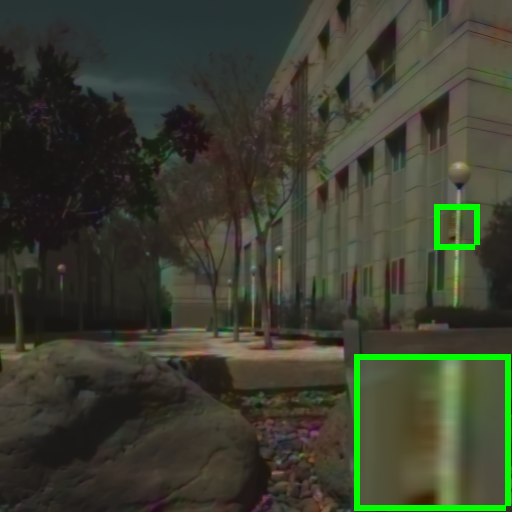}\\(e) DeSCI
 \end{minipage}\\

 \begin{minipage}[t]{0.18\textwidth}\centering
   \includegraphics[width=\textwidth]{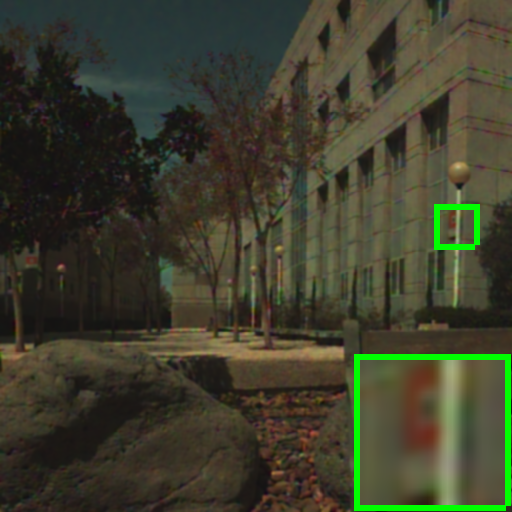}\\(f) TV-RGB
 \end{minipage}
 \begin{minipage}[t]{0.18\textwidth}\centering
   \includegraphics[width=\textwidth]{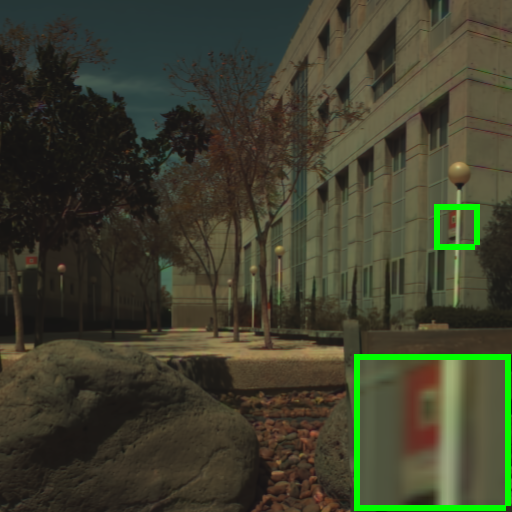}\\(g) DeSCI-RGB
 \end{minipage}
 \begin{minipage}[t]{0.18\textwidth}\centering
   \includegraphics[width=\textwidth]{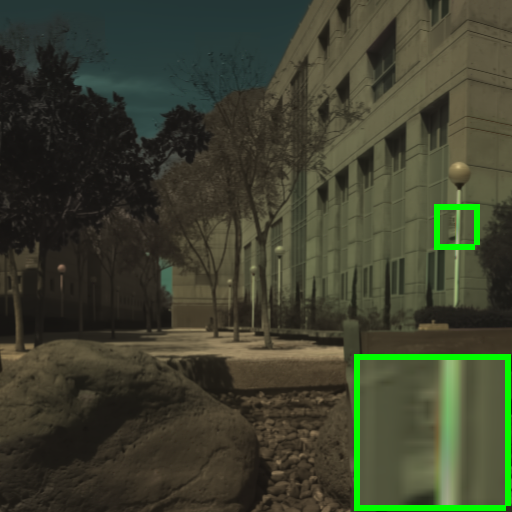}\\(h) DLTR
 \end{minipage}
 \begin{minipage}[t]{0.18\textwidth}\centering
   \includegraphics[width=\textwidth]{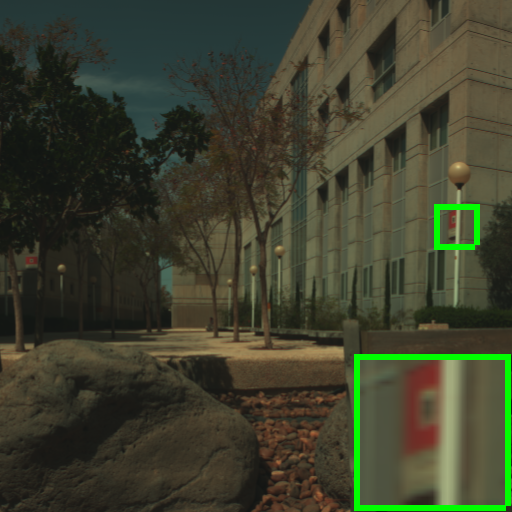}\\(i) Fusion
 \end{minipage}
 \begin{minipage}[t]{0.18\textwidth}\centering
   \includegraphics[width=\textwidth]{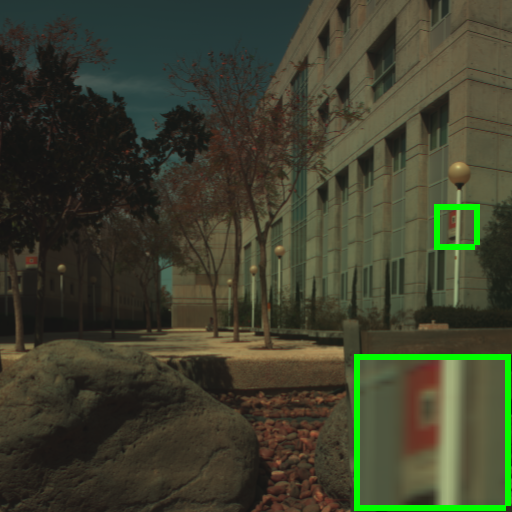}\\(j) PFusion
 \end{minipage}\\
\caption{Reconstructed results of different methods on ICVL-BGU image. The color images are composed of bands 31, 11, 6 for red, green, and blue, respectively.}
\label{fig:ICVL_BGU}
 \end{figure*}

\begin{figure}[!htp]
 \centering
 \begin{minipage}[t]{0.24\textwidth}\centering
   \includegraphics[width=\textwidth]{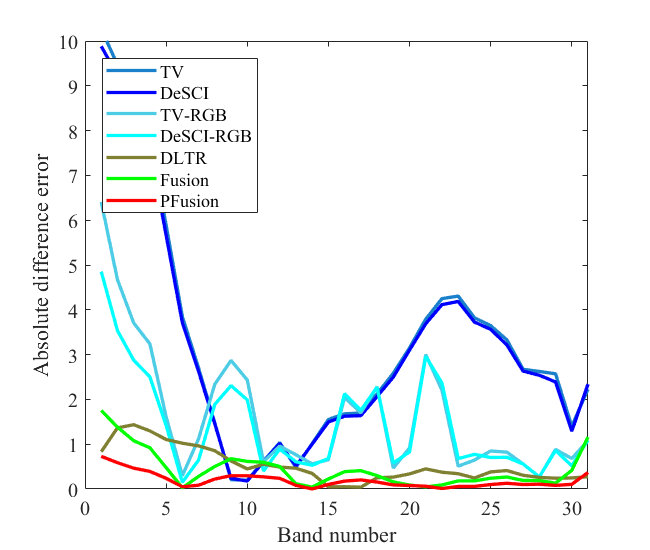}\\(a) BGU
 \end{minipage}
 \begin{minipage}[t]{0.24\textwidth}\centering
   \includegraphics[width=\textwidth]{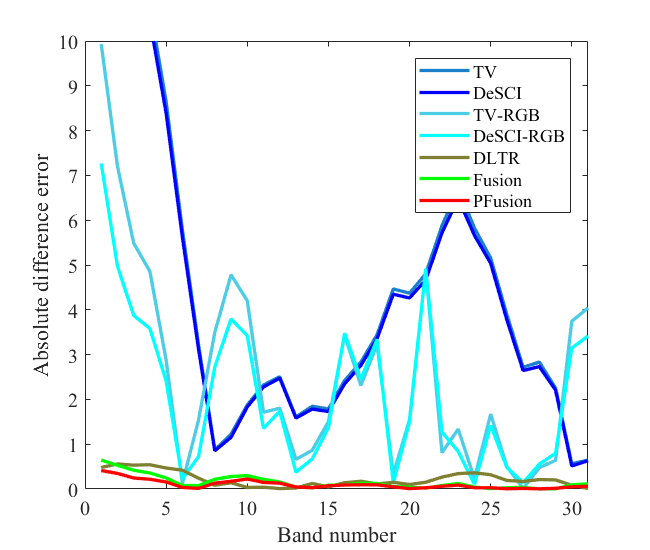}\\(b) IDS
 \end{minipage}\\
 \caption{The absolute difference errors between the mean of the whole original signatures and the one obtained by different reconstruction methods on ICVL images.}
 \label{fig:ICVL_sig}
\end{figure}

\begin{table}[htbp]
  \centering
  \caption{Quantitative evaluation of EO-1 data experiments for different HSI reconstruction methods from CASSI and RGB measurements.}
  \setlength{\tabcolsep}{1.1mm}{
    \begin{tabular}{cccccccc}
    \toprule
    Method    & TV    & DeSCI & TV-RGB & DeSCI-RGB & DLTR  & Fusion & PFusion \\
    \midrule
    M-PSNR & 26.64 & 28.96 & 29.91 & 37.01 & 35.56 & 41.8  & \textbf{43.34} \\
    M-SSIM & 0.839 & 0.948 & 0.931 & 0.977 & 0.963 & 0.983 & \textbf{0.984} \\
    MSA   & 4.929 & 4.253 & 3.864 & 1.6   & 1.78  & 1.03  & \textbf{0.85} \\
    \bottomrule
    \end{tabular}%
  \label{tab:eva3}
  }%
\end{table}%
\begin{figure*}[!htp]
  \scriptsize
  \centering
 \begin{minipage}[t]{0.09\textwidth}\centering
   \includegraphics[width=\textwidth]{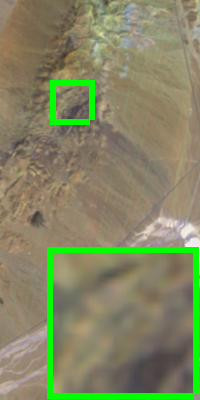}\\(a) Original
 \end{minipage}
 \begin{minipage}[t]{0.09\textwidth}\centering
   \includegraphics[width=\textwidth]{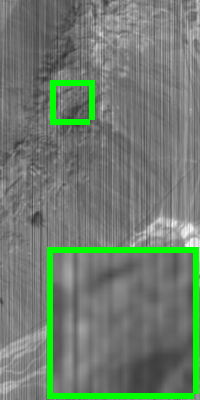}\\(b) CASSI
 \end{minipage}
 \begin{minipage}[t]{0.09\textwidth}\centering
   \includegraphics[width=\textwidth]{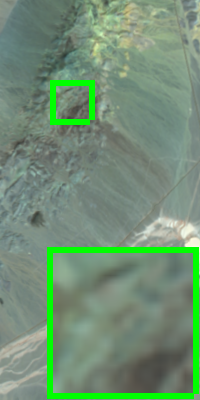}\\(c) RGB
 \end{minipage}
 \begin{minipage}[t]{0.09\textwidth}\centering
   \includegraphics[width=\textwidth]{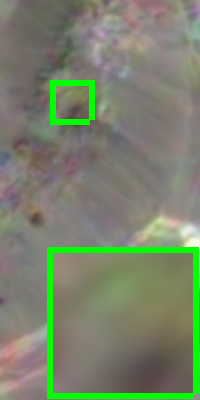}\\(d) TV
 \end{minipage}
 \begin{minipage}[t]{0.09\textwidth}\centering
   \includegraphics[width=\textwidth]{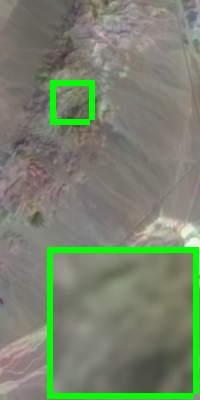}\\(e) DeSCI
 \end{minipage}
 \begin{minipage}[t]{0.09\textwidth}\centering
   \includegraphics[width=\textwidth]{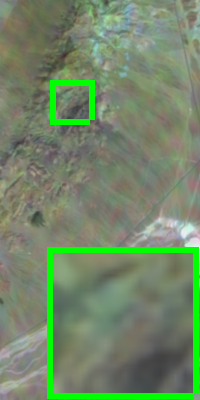}\\(f) TV-RGB
 \end{minipage}
 \begin{minipage}[t]{0.09\textwidth}\centering
   \includegraphics[width=\textwidth]{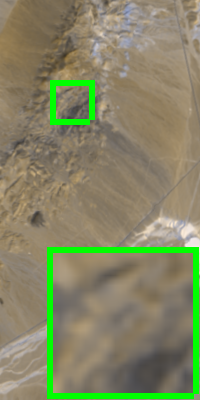}\\(g) DeSCI-RGB
 \end{minipage}
 \begin{minipage}[t]{0.09\textwidth}\centering
   \includegraphics[width=\textwidth]{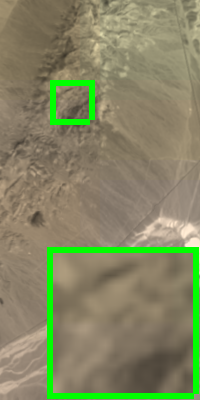}\\(h) DLTR
 \end{minipage}
 \begin{minipage}[t]{0.09\textwidth}\centering
   \includegraphics[width=\textwidth]{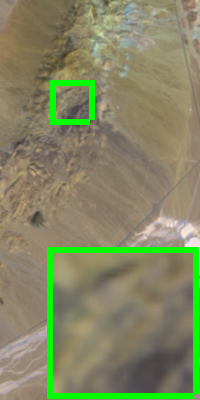}\\(i) Fusion
 \end{minipage}
 \begin{minipage}[t]{0.09\textwidth}\centering
   \includegraphics[width=\textwidth]{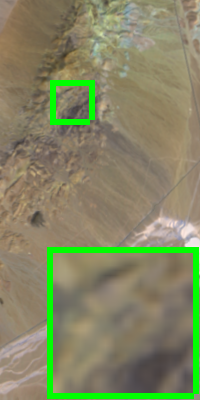}\\(j) PFusion
 \end{minipage}\\
\caption{Reconstructed results of different methods on EO-1 image. The color images are composed of bands 100, 60, 10 for red, green, and blue, respectively.}
\label{fig:RS}
 \end{figure*}

\begin{figure*}[!htp]
 \centering
 \begin{minipage}[t]{0.30\textwidth}\centering
   \includegraphics[width=\textwidth]{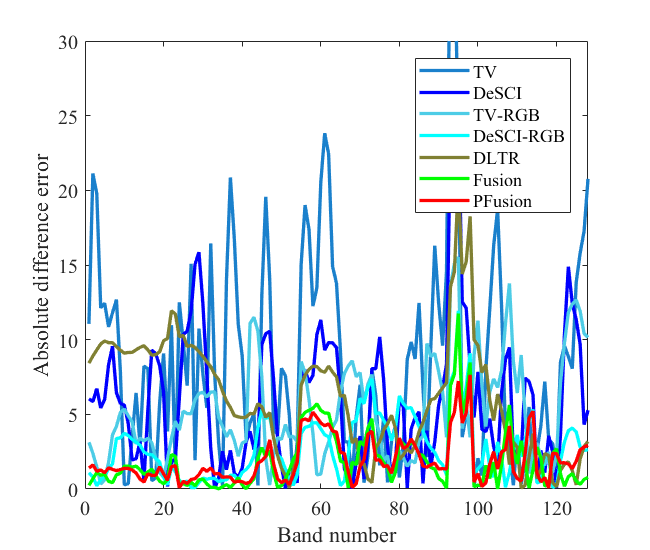}\\(a) point 1
 \end{minipage}
 \begin{minipage}[t]{0.30\textwidth}\centering
   \includegraphics[width=\textwidth]{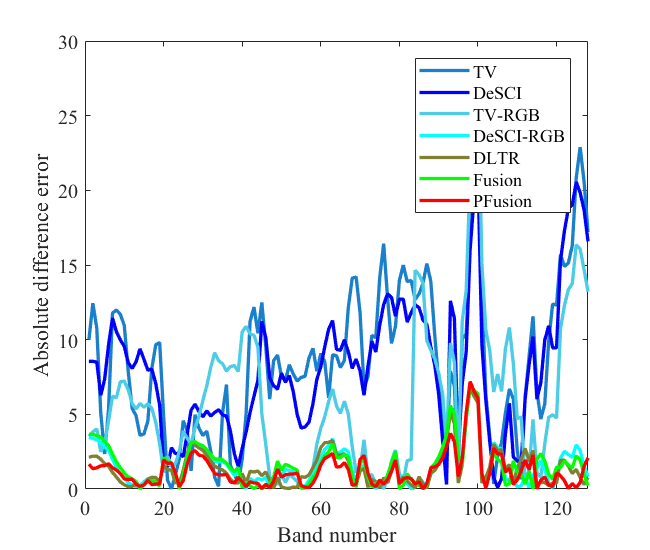}\\(b) point 2
 \end{minipage}
 \begin{minipage}[t]{0.30\textwidth}\centering
   \includegraphics[width=\textwidth]{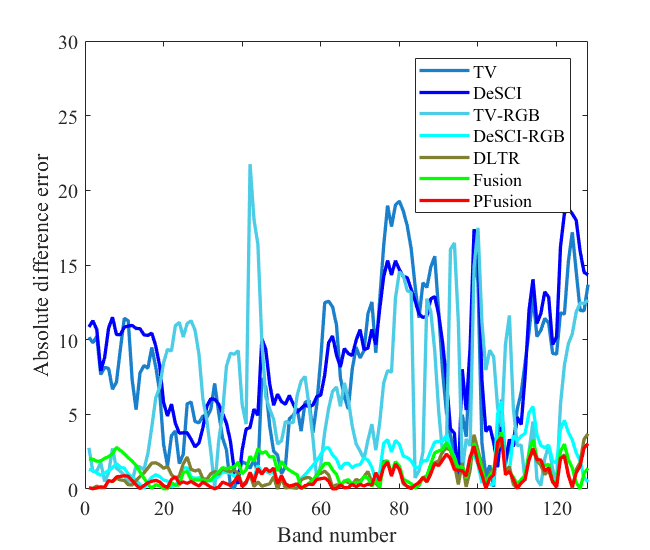}\\(c) point 3
 \end{minipage}\\
 \caption{The absolute difference errors between the original signature and the one obtained by different reconstruction methods on EO-1 image.}
 \label{fig:RS_sig}
\end{figure*}

\begin{figure*}[htp]
  \footnotesize
  \centering
 \begin{minipage}[t]{0.16\textwidth}\centering
   \includegraphics[width=\textwidth]{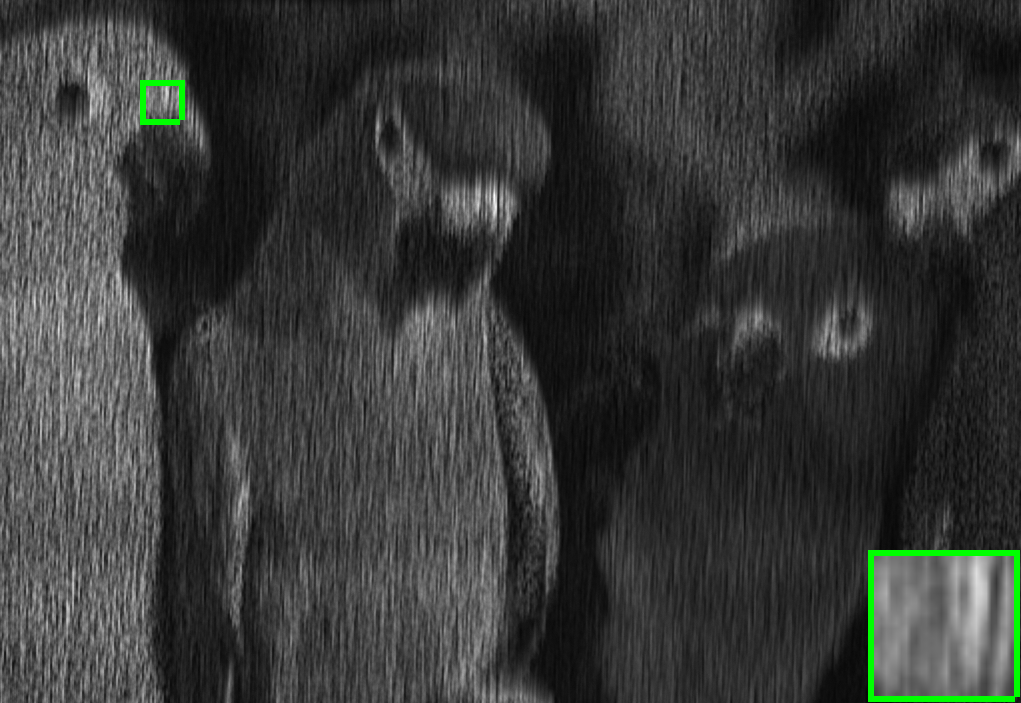}\\(a) CASSI
 \end{minipage}
 \begin{minipage}[t]{0.16\textwidth}\centering
   \includegraphics[width=\textwidth]{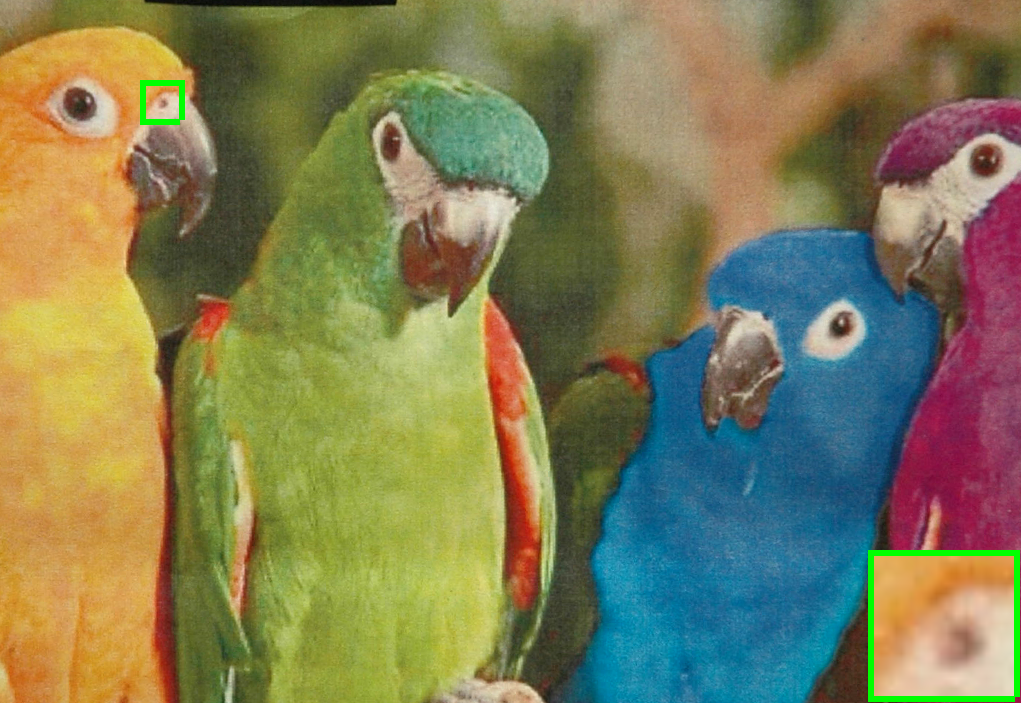}\\(b) RGB
 \end{minipage}
 \begin{minipage}[t]{0.16\textwidth}\centering
   \includegraphics[width=\textwidth]{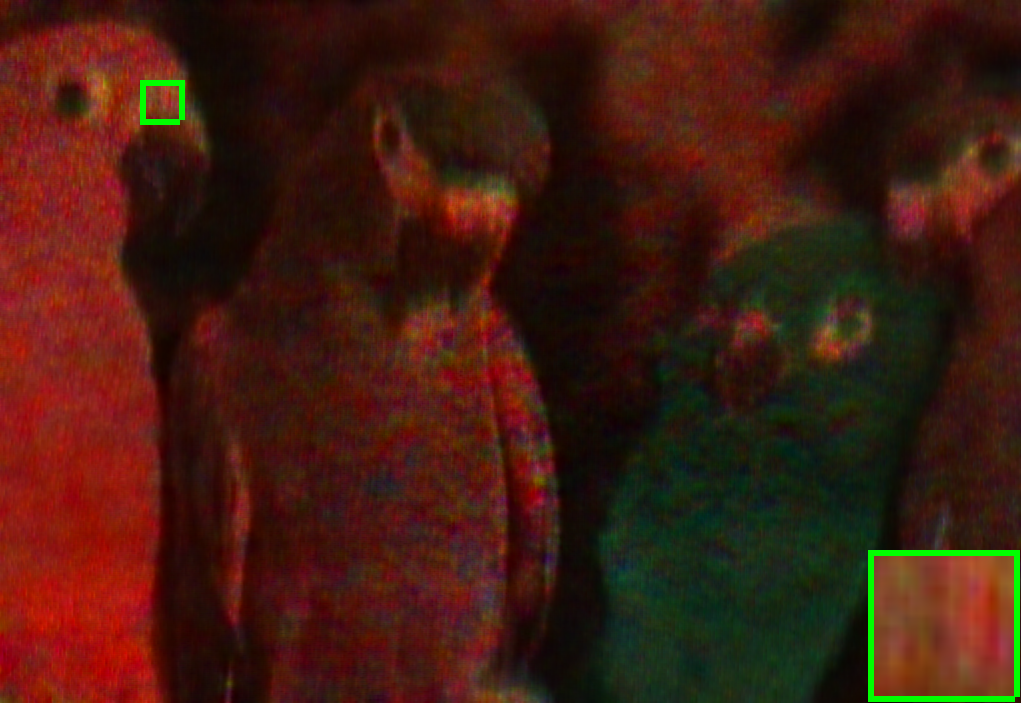}\\(c) TV
 \end{minipage}
 \begin{minipage}[t]{0.16\textwidth}\centering
   \includegraphics[width=\textwidth]{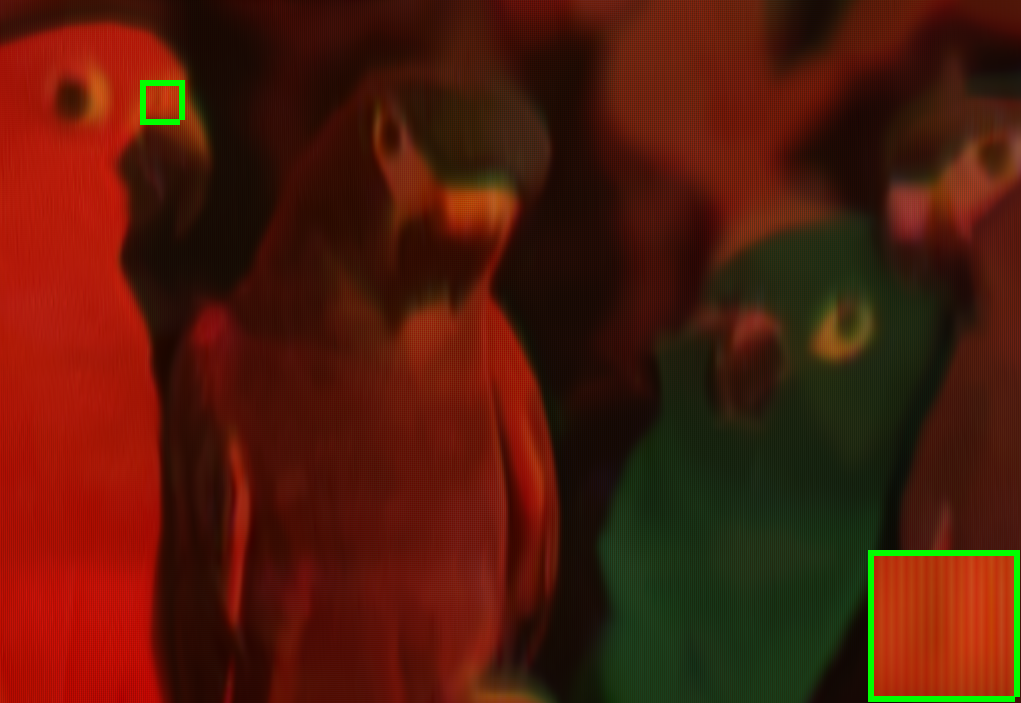}\\(d) DeSCI
 \end{minipage}
 \begin{minipage}[t]{0.16\textwidth}\centering
   \includegraphics[width=\textwidth]{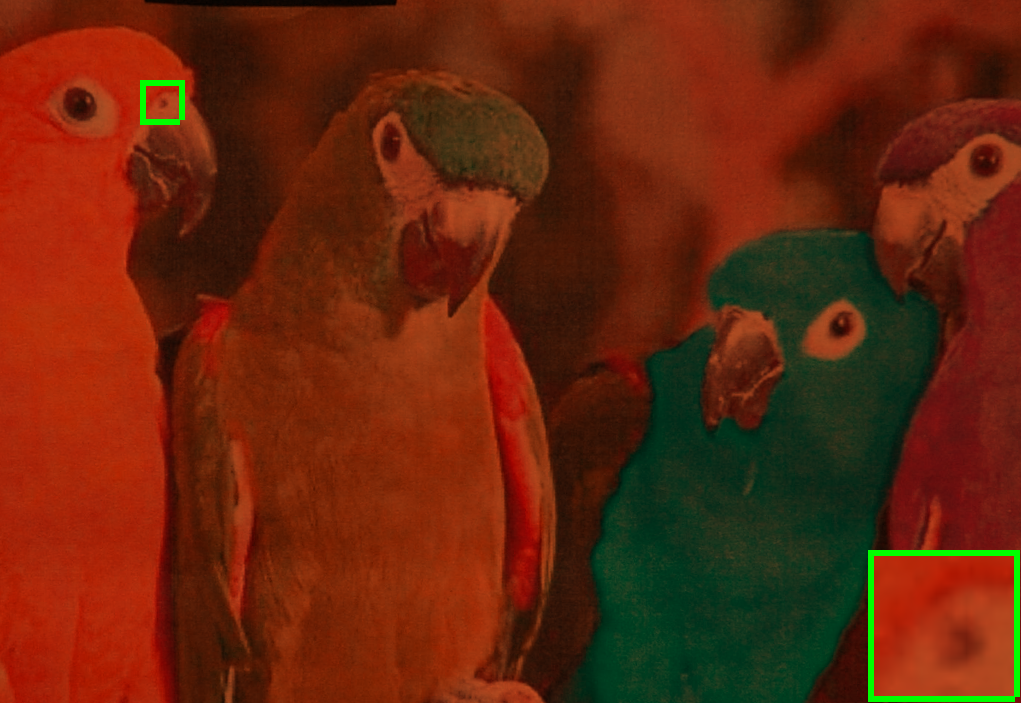}\\(e) Fusion
 \end{minipage}
 \begin{minipage}[t]{0.16\textwidth}\centering
   \includegraphics[width=\textwidth]{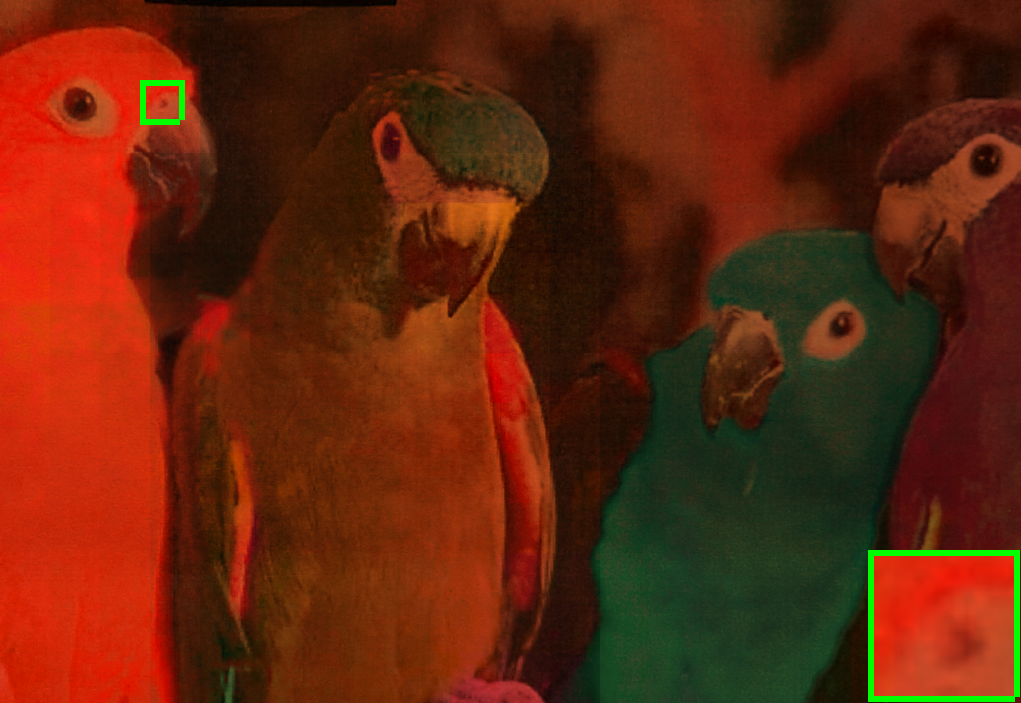}\\(f) PFusion
 \end{minipage}\\
\caption{Reconstructed results of different methods on real Bird image. The color images are composed of bands 20, 10, 6 for red, green, and blue, respectively.}
\label{fig:bird}
 \end{figure*}

\begin{figure*}[htp]
 \centering
 \begin{minipage}[t]{0.30\textwidth}\centering
   \includegraphics[width=\textwidth]{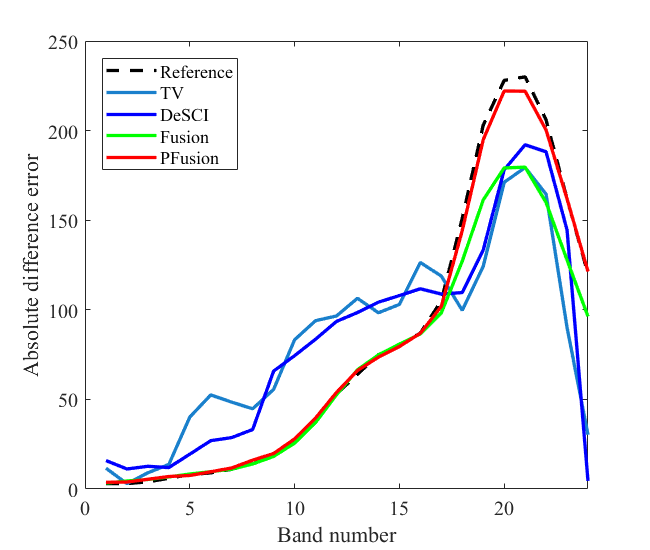}\\(a) (100,100)
 \end{minipage}
 \begin{minipage}[t]{0.30\textwidth}\centering
   \includegraphics[width=\textwidth]{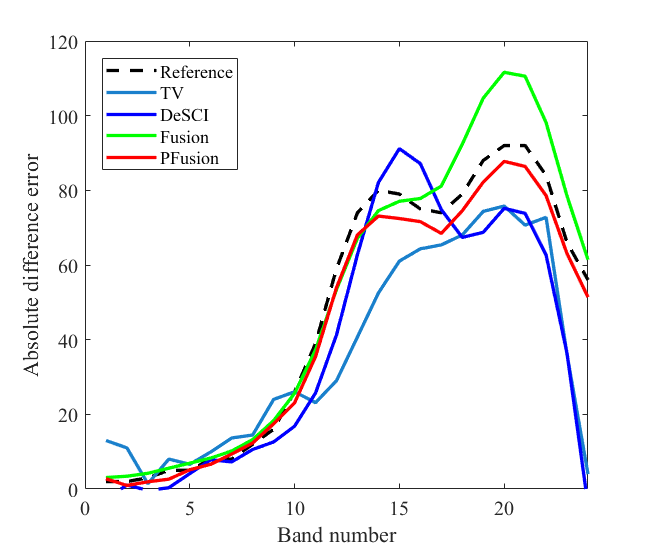}\\(b) (350,350)
 \end{minipage}
 \begin{minipage}[t]{0.30\textwidth}\centering
   \includegraphics[width=\textwidth]{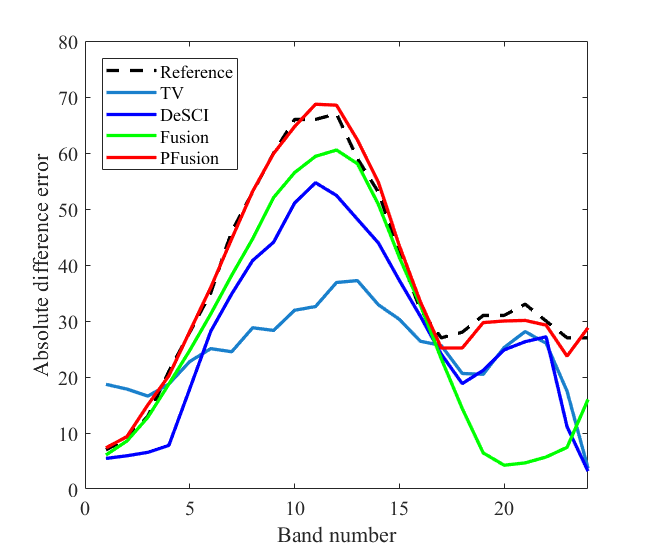}\\(c) (700,700)
 \end{minipage}\\
 \caption{The absolute difference errors between the original signature and the one obtained by different reconstruction methods on real Bird image.}
 \label{fig:bird_sig}
\end{figure*}

The second test dataset is ICVL\footnote{\url{http://icvl.cs.bgu.ac.il/hyperspectral/}}. We simulate the CASSI measurements and RGB measurements the same as that of CAVE dataset. We select $5$ images from ICVL dataset for the experiments. The original image from ICVL is of size $1392 \times 1304 \times 31$ and we reshape it to the size of $512 \times 512 \times 31$. For the proposed methods, we still choose $k=3$ and $m=n=100$.

\noindent
\textbf{Quantitative comparison.}
Table \ref{tab:eva2} presents the quantitative comparison results of reconstructed images via different methods on $5$ ICVL images. Again, the reconstruction methods from only the CASSI measurements $i.e.,$ TV and DeSCI achieve the worst values of M-PSNR, M-SSIM and MSA. The combined measurements of CASSI and RGB can significantly improve the reconstruction results, just as TV-RGB. DeSCI-RGB and DLTR further improve the results. DeSCI-RGB can achieve higher M-PSNR and M-SSIM values, meanwhile DLTR obtains higher MSA values. Our proposed PFusion significantly improves the spectral quality compared to Fusion, and again achieves the best quantitative values across different ICVL images, further indicating the advantage of our proposed strategy to process the reconstruction of CASSI and RGB measurements.

\noindent
\textbf{Visual comparison.}
Fig.~\ref{fig:ICVL_BGU} presents the color images (composed of bands 31, 11 and 6) of different reconstruction methods on ICVL BGU image. It can be clearly observed that RGB measurements can provide the spatial structure information, but totally different spectral information. The reconstruction result of TV from the CASSI measurements is blurred. DeSCI and TV-RGB can improve the results, but they are still blurred. DeSCI-RGB and DLTR further improve the results. However, as presented in the enlarged rectangle of Fig.~\ref{fig:ICVL_BGU}(h), DLTR loses the color of the board. To sum up, the proposed Fusion and PFusion produce the best results, with much more detailed information compared to the original image. Fig.~\ref{fig:ICVL_sig} illustrates the absolute values between the mean of the original signatures and the one obtained by different reconstruction methods of $3$ signatures from three different ICVL images. Compared to DeSCI-RGB and DLTR, the advantage of the proposed Fusion is limited to preserve the spectral information. However, the proposed PFusion can produce the smallest absolute difference errors, indicating the best performance in the preservation of HSI spectrum.

\subsection{Simulated Experiments On Remote Sensing}
We further conduct the remote sensing compressive reconstruction experiments to testify the efficiency of the proposed methods. The HSI is from the earth-observing One (EO-1) Hyperion sensor~\footnote{\url{https://archive.usgs.gov/archive/sites/eo1.usgs.gov/index.html}}.
Similar to \cite{Yuan_PAMI_2019}, we adopt CASSI to compress the HSI, meanwhile the spectral sensing matrix to obtain RGB measurements. The size of HSI, CASSI measurements and RGB measurements are presented in Table \ref{tab:datasize}.

\noindent
\textbf{Quantitative comparison.}
We adopt Table \ref{tab:eva3} to present the quantitative comparison results of reconstructed images via different methods on the EO-1 image. From the table, it can be observed that the proposed PFusion obtains the best accuracy in three different metrics. TV-RGB, DeSCI-RGB and DLTR achieves better results than that of TV and DeSCI, indicating that the reconstruction results from the combined CASSI and RGB measurements are better those of only CASSI measurements.

\noindent
\textbf{Visual comparison.}
Fig.~\ref{fig:RS} presents the color images (composed of bands 100, 60 and 10) of different reconstruction methods on the EO-1 image. We can observe that our proposed Fusion and PFusion outperform others to preserve the details of the reconstructed image. Other comparison methods produce the blurred results. Fig.~\ref{fig:RS_sig} shows the absolute difference errors between the original signature and the ones obtained by different reconstruction methods on the EO-1 image. Our proposed PFusion again beats other comparison methods, and achieves the smallest absolute difference errors. That is to say, our proposed Fusion and PFusion can preserve the spatial details, meanwhile, the overlapping patches processing of PFusion can significantly improve the capability to preserve the spectral information.

\subsection{Real Experiments On Bird}
In this section, we apply the proposed method to the HSI reconstruction of the real snapshot-hyperspectral compressive imaging data. The real Bird data is from CASSI system~\cite{CASSI2008}, which has been widely used for the real experiments analysis~\cite{Yuan_PAMI_2019,side_2015}. The size of reconstructed Bird image, CASSI and RGB measurements are presented in Table~\ref{tab:datasize}. Since we do not know the spectral sensing matrix from HSI to RGB measurements, we cannot implement TV-RGB, DeSCI-RGB and DLTR for comparison. So we simply compare our proposed methods to TV and DeSCI\footnote{The results of TV and DeSCI are provided by Dr. Y. Liu at website https://github.com/liuyang12/DeSCI}. \par

\noindent
\textbf{Visual comparison.}
Fig.~\ref{fig:bird} illustrates the color images (composed of bands 20, 10 and 6) of different reconstruction methods on the real bird data. We may find that the measured RGB image contains abundance spatial information. However, the reconstructed images via TV and DeSCI lose the spatial details. By contrast, our proposed Fusion and PFusion are capable of preserving the spatial information from RGB measurements. Fig.~\ref{fig:bird_sig} presents the signatures of three points obtained by different methods. It can be observed that our proposed PFusion are closest to the reference signatures. It indicates that our proposed PFusion can preserve the spectral information, therefore, again demonstrating the advantage of our proposed method.

\section{Ablation Study and Discussion}

\subsection{Spectral Sensing Matrix Analysis}
\begin{figure}[!htp]
\footnotesize
 \centering
 \begin{minipage}[t]{0.45\textwidth}\centering
   \includegraphics[width=\textwidth]{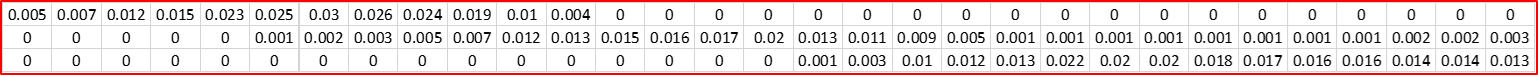}(a)
 \end{minipage}
 \begin{minipage}[t]{0.45\textwidth}\centering
   \includegraphics[width=\textwidth]{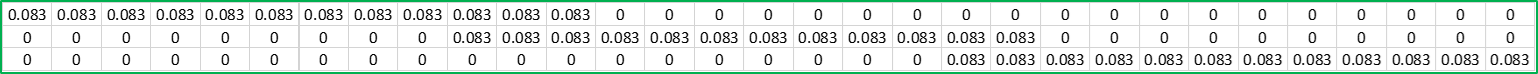}(b)
 \end{minipage}
 \begin{minipage}[t]{0.45\textwidth}\centering
   \includegraphics[width=\textwidth]{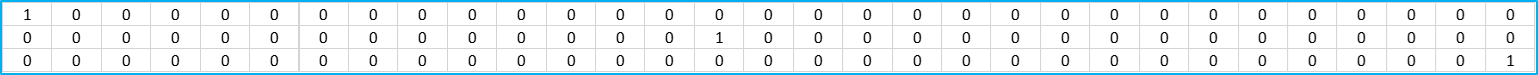}(c)
 \end{minipage}\\
 \caption{Different spectral sensing matrix to measure RGB from HSI. (a) Provided by Nikon D700 camera, (b) average based design, and (c) single band selection.}
 \label{fig:sensor}
\end{figure}

We first analyze the proposed methods on different spectral sensing matrices. The spectral sensing matrix of RGB detector is adopted to measure the RGB image from HSI, and the RGB measurement is adopted to provide the coefficients, as illustrated in Fig.~\ref{fig:fuse}. Fig.~\ref{fig:sensor} presents different kinds of designed spectral sensing matrices, including (a) provided by Nikon D700 camera, (b) average of the bands, and (c) random selection of three single bands. Table~\ref{tab:sensor} shows the average quantitative evaluation results of proposed methods with different sensor matrices on 10 CAVE HSIs. It can be clearly observed that the proposed Fusion and PFusion can produce almost similar results with spectral sensing matrices provided by Nikon D700 camera and average design strategy. On the other hand, the random single band selection strategy produces the worse results. This comparison of different spectral sensing matrices guides us to design the RGB measurements efficiently. This suggests that the RGB measurement should try to cover all bands information. In this case, the coefficients estimated from the measured RGB image will provide more useful information for the subsequent HSI reconstruction.

\begin{table}[htbp]
  \centering
  \caption{Quantitative evaluation of CAVE data experiments for proposed methods with different sensor matrices.}
    \begin{tabular}{c|ccc|ccc}
    \toprule
    \multirow{2}[2]{*}{sensing matrix} & \multicolumn{3}{c|}{Fusion} & \multicolumn{3}{c}{PFusion} \\
          & (a)   & (b)   & (c)   & (a)   & (b)   & (c) \\
    \midrule
    MPSNR & \textbf{38.97} & 38.47 & 35.32 & \textbf{43.42} & 43.39 & 37.76 \\
    MSSIM & \textbf{0.962} & 0.955 & 0.906 & \textbf{0.984} & \textbf{0.984} & 0.955 \\
    MSA   & 17.25 & \textbf{16.06} & 16.86 & 7.39  & \textbf{6.16} & 7.61 \\
    \bottomrule
    \end{tabular}%
  \label{tab:sensor}%
\end{table}%
\subsection{Rank $k$ Analysis}
\begin{figure}[!htp]
 \centering
 \begin{minipage}[t]{0.15\textwidth}\centering
   \includegraphics[width=\textwidth]{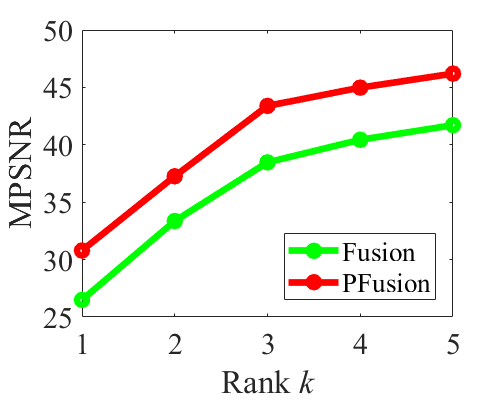}\\(a) MPSNR
 \end{minipage}
 \begin{minipage}[t]{0.15\textwidth}\centering
   \includegraphics[width=\textwidth]{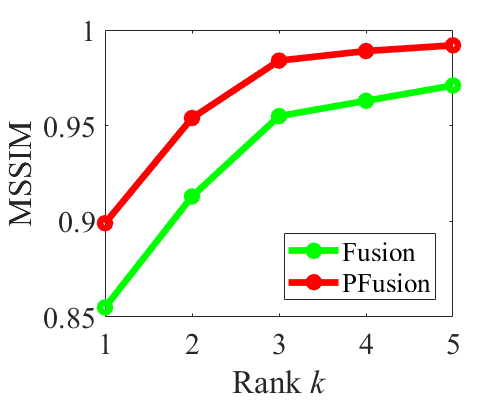}\\(b) MSSIM
 \end{minipage}
 \begin{minipage}[t]{0.15\textwidth}\centering
   \includegraphics[width=\textwidth]{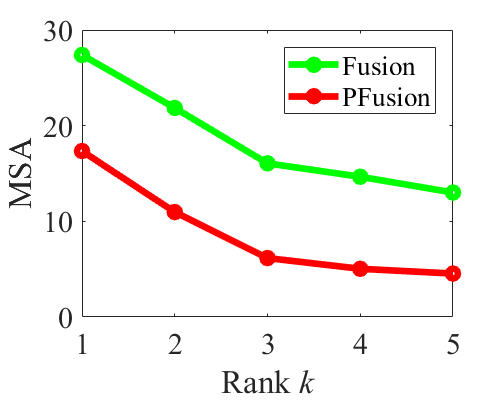}\\(c) MSA
 \end{minipage}\\
 \caption{Changes of quantitative evaluation results obtained by the proposed methods with different rank $k$. We test on 10 CAVE HSIs.}
 \label{fig:dimen}
\end{figure}

Subsequently, we analyze the influence of rank $k$. Note that, we assume that HSIs lie in an approximate low-dimensional spectral subspace, and therefore, a larger rank $k$ will assure more precise low-rank approximation for the original HSI as presented in~\eqref{eq:low-rank}. However, in our fusion model~\eqref{eq:fusion1} and~\eqref{eq:fusion11}, the rank $k$ is bounded by the size of spectral sensing matrix $\mat{A}$. In the proposed methods, we set $k=3$ (3 is the column size of $\mat{A}$) to explore more information of HSI. In this section, we change the size of spectral sensing matrix $\mat{A}$, and the measured RGB image is changed from 1-band panchromatic image to 5-band multispectral image, and therein the rank $k$ changes from 1 to 5. Fig.~\ref{fig:dimen} presents the quantitative evaluation results obtained by the proposed methods with different rank $k$. As illustrated, with the increase of rank $k$, the quantitative evaluation results increase significantly. This is reasonable, since more measurements suggest higher reconstruction accuracy. From the figure, we can observe that the complementary 3-band RGB measurements can produce much higher HSI reconstruction results compared to complementary panchromatic measurements. By contrast, the improvement from 3-band RGB to multispectral measurements with more than 3-band is limited. What's more, the multispectral measurements will bring huge measured burden for the hardware. Therefore, to realize our proposed methods on the real hardware, we recommend the complementary 3-band RGB measurements.

\subsection{Patch Spatial Size Analysis}
\begin{table}[!htbp]
  \centering
  \caption{Changes of quantitative evaluation results obtained by the proposed PFusion with patch size $m$. We test on on 10 CAVE HSIs.}
  \setlength{\tabcolsep}{1.7mm}{
    \begin{tabular}{cccccccc}
    \toprule
    patch size & 20    & 40    & 60    & 100   & 160   & 320   & 512 \\
    \midrule
    MPSNR & 41.72 & 43.17 & 43.41 & 43.42 & 42.31 & 40.03 & 38.97 \\
    MSSIM & 0.988 & 0.989 & 0.985 & 0.984 & 0.979 & 0.967 & 0.962 \\
    MSA   & 6.72  & 6.98  & 7.29  & 7.39  & 9.94  & 15.82 & 17.25 \\
    \bottomrule
    \end{tabular}%
  \label{tab:patchsize}%
  }
\end{table}%

As analyzed in Section~\ref{Analysis1} and the experimental results presented in Section~\ref{Experiments}, we can conclude that the proposed PFusion with patch processing can significantly improve the reconstruction results compared to Fusion. For our proposed PFusion on the 10 CAVE images, we change the patch size $m$ (we set $m=n$) from $20$ to $512$, and present the quantitative evaluation results in table~\ref{tab:patchsize}. It can be observed that our proposed PFusion is robust when the patch size changes from $40$ to $100$. On the other hand, when the patch size becomes larger than $320$, the performance decreases significantly. This is mainly because the rank $k=3$ is not enough to describe the spectral information of such big patches. Meanwhile, the smaller patch size will decrease the quality of orthogonal spectral basis estimation, as analyzed in Section~\ref{Analysis1}. It should be noticed that the patch size should be also related to the complexity of the image. If the image is complex and contains multiple materials, the patch size should be smaller, and vice versa. In the whole experiments, we fix the patch size as 100.

\subsection{Improved Spectral Basis Estimation}
\label{sec:improvel}
We present an improved orthogonal spectral basis estimation method in \eqref{eq:fusion22_ext}. Compared to original \eqref{eq:fusion22}, the improved  \eqref{eq:fusion22_ext} combines CASSI and RGB measurements to estimate the orthogonal spectral basis. Table \ref{tab:ext} illustrates the quantitative evaluation results obtained by PFusion and the improved method \eqref{eq:fusion22_ext} on different datasets. We can clearly observe that the improved method can slightly ($\le$0.16dB in M-PSNR) enhance the results. However, it brings more than three times additional computational burden compared to the original case \eqref{eq:fusion22}. It is in fact a trade-off between the accuracy and speed of the HSI reconstruction.

\begin{table}[htbp]
  \centering
  \caption{Quantitative evaluation results obtained by PFusion and the improved on different dataset.}
  \setlength{\tabcolsep}{1.1mm}{
    \begin{tabular}{c|cc|cc|cc}
    \toprule
    \multirow{2}[2]{*}{} & \multicolumn{2}{c|}{CAVE} & \multicolumn{2}{c|}{ICVL} & \multicolumn{2}{c}{RS} \\
          & PFusion & Improved & PFusion & Improved & PFusion & Improved \\
    \midrule
    M-PSNR & 43.42 & \textbf{43.46} & 46.91 & \textbf{47.07} & 43.34 & \textbf{43.38} \\
    M-SSIM & \textbf{0.984} & \textbf{0.984} & \textbf{0.995} & \textbf{0.995} & 0.984 & \textbf{0.985} \\
    MSA   & 7.39  & \textbf{7.27} & \textbf{1.15} & 1.16  & 0.85  & \textbf{0.81} \\
    Time  & \textbf{1.8} & 9.2   & \textbf{1.8} & 9.2   & \textbf{5.1} & 18.2 \\
    \bottomrule
    \end{tabular}%
  \label{tab:ext}%
  \vspace{-5px}
  }
\end{table}%

\subsection{Computational Efficiency}
\begin{table}[htbp]
  \centering
  \caption{Running time (in seconds) of different methods on different dataset.}
  \setlength{\tabcolsep}{1.1mm}{
    \begin{tabular}{cccccccc}
    \toprule
    Method & TV    & DeSCI & TV-RGB & DeSCI-RGB & DLTR  & Fusion & PFusion \\
    \midrule
    CAVE  & 251   & 3787  & 523   & 12544  & 52585 & 1.3   & 1.8 \\
    ICVL  & 250   & 3776  & 511   & 12549  & 48572 & 1.3   & 1.8 \\
    RS    & 1123  & 8288  & 1478  & 18297  &  79234 & 3.3  & 5.1 \\
    Bird  & 472   & 18690 & *     & *      & *     & 2.8   & 3.9 \\
    \bottomrule
    \end{tabular}%
  \label{tab:time}%
  }
\end{table}%
Finally, we analyze the computational efficiency of the proposed methods. The cost time of TV and DeSCI methods on real Bird image are provided by \cite{Yuan_PAMI_2019}. Table \ref{tab:time} presents the running time of different methods on different dataset. TV-RGB and DeSCI-RGB cost much more time compared to TV and DeSCI due to the additional measurements \eqref{eq:CASSI_RGB_pre}. Non-local low-rank related methods DeSCI-RGB and DLTR~\footnote{The cost time was provided by Dr. S. Zhang} can obtain the satisfied reconstruction results, however the computational burden is also huge. In particular, our proposed methods can reconstruct the HSI within several seconds, more than 5000 times faster than that of non-local related methods DeSCI-RGB and DLTR. This demonstrates a huge advantage of our proposed model to decompose the full HSI into orthogonal spectral basis and spatial coefficients, and then reconstruct the two smaller components separately. Compared to Fusion, PFusion needs additional computational time. However, PFusion significantly improves the reconstruction accuracy, as presented in Table~\ref{tab:patchsize}. To conclude, our proposed model~\eqref{eq:fusion1},~\eqref{eq:fusion2} is proved to be efficient to obtain the best accuracy, meanwhile much less computational time.

\section{CONCLUSION}
In this study, we have proposed a new model to reconstruct the HSI from CASSI and RGB measurements. We explore the spectral low-rank property of HSI, and decompose it to the orthogonal spectral basis and spatial coefficients. The RGB measurements can provide the estimation of coefficients, meanwhile CASSI measurements are adopted to estimate the orthogonal spectral basis. Compared to the previous works that try to reconstruct full HSIs with different regularizers, our proposed Fusion and PFusion methods do not require non-local processing or iteration, which can save huge computational time. Furthermore, our proposed methods does not require the spectral sensing matrix in advance. The experiments on three simulated HSI datasets and one real dataset demonstrated that our proposed Fusion and PFusion can reconstruct the HSI with the highest accuracy using far less computational time. In summary, our proposed methods can reconstruct the HSIs in a fast, flexible and high accuracy manner. As illustrated in \eqref{eq:general}, we did not add any additional regularizer on our model, to alleviate the computational burden. However, as presented in Section~\ref{Analysis1}, our model still has room for improvement to meet the condition of $k\leqslant 3$. We hope that the proposed optimization can be used as a baseline, and regard the improvement of our proposed model as future work.

\section{Appendix}

\subsection{Proposition 1}
\begin{proof}
The proof can be decomposed to two steps. First, the original HSI can be uniquely decomposed to $\mat{X} = \mat{E} \mat{W}$, where $\mat{W}$ is computed via SVD on $\mat{Z}$. Using SVD, we have $\mat{Z} = (\mat{A}^{\top}\mat{X}) = \mat{F} \mat{W}$, where $\mat{F} \in \mathbb{R}^{3 \times K}$. Since $k\leqslant 3$ and $\mat{A}$ is of full-column rank, we can conclude that $\mat{F}$ is of full-column rank, and $\mat{W}$ is of full-row rank. If we have another $\mat{E}^{'}$ such that $\mat{X} = \mat{E}^{'} \mat{W}$. We can obtain $(\mat{E}-\mat{E}^{'}) \mat{W}=\mat{0}$. Since $\mat{W}$ is of full-row rank, we can conclude that $\mat{E}=\mat{E}^{'}$. Therefore, the decomposition $\mat{X} = \mat{E} \mat{W}$ is unique. \par
Second, $\mat{E}$ can be uniquely obtained via \eqref{eq:opt22}. We have obtained the full-row rank $\mat{W}$ via SVD on $\mat{Z}$. Since the full-row rank property of $\mat{W}$, and the specific design of CASSI mask $\tensor{C}$ as in \eqref{eq:phi}, we have the condition that $\mat{\Phi}^{\mat{W}}$ is of full-column rank. Therefore, the solution of \eqref{eq:fusion22} is an overdetermined equation problem. That is to say, the obtained $\mat{E}$ from \eqref{eq:fusion22} is unique.

Thus, we obtain the proposition.

\end{proof}
\ifCLASSOPTIONcaptionsoff
  \newpage
\fi

\bibliographystyle{IEEEtran}
\bibliography{IEEEabrv,lowrank_review_references}

\vfill
\end{document}